\documentclass[11pt]{article}
\usepackage[margin=1in]{geometry}
\usepackage{authblk}
\usepackage{amsthm,amsmath,amsfonts,amssymb}
\usepackage{adjustbox}
\usepackage{subcaption}

\usepackage{algorithm}%
\usepackage{algpseudocode}

\usepackage[numbers]{natbib}
\usepackage[colorlinks,citecolor=blue,urlcolor=blue]{hyperref}
\usepackage{graphicx}

\usepackage{lipsum}
\usepackage{epstopdf}
\usepackage{nameref}
\usepackage{tikz}
\usetikzlibrary{arrows.meta,calc,tikzmark,decorations.pathmorphing}

\ifpdf
  \DeclareGraphicsExtensions{.eps,.pdf,.png,.jpg}
\else
  \DeclareGraphicsExtensions{.eps}
\fi

\usepackage{enumitem}
\setlist[enumerate]{leftmargin=.5in}
\setlist[itemize]{leftmargin=.5in}

\theoremstyle{plain}

\newtheorem{theorem}{Theorem}[section]
\newtheorem{lemma}[theorem]{Lemma}
\newtheorem{model}{Model}
\newtheorem{property}{Property}
\newtheorem{corollary}{Corollary}

\newcommand{\EE}{\mathbb{E}}
\newcommand{\supp}{\operatorname{supp}}
\newcommand{\R}{\mathbb{R}}
\newcommand{\RR}{\mathbb{R}}
\newcommand{\PP}{\mathbb{P}}
\newcommand{\psim}{\psi^{(m)}}
\newcommand{\tpsim}{\tilde{\psi}^{(m)}}
\newcommand{\hpsim}{\hat{\psi}^{(m)}}

\newcommand{\bA}{\mathbf{A}}

\newcommand{\bD}{\mathbf{D}}

\newcommand{\bP}{\mathbf{P}}

\newcommand{\bS}{\mathbf{S}}

\newcommand{\bU}{\mathbf{U}}

\newcommand{\bW}{\mathbf{W}}
\newcommand{\bX}{\mathbf{X}}

\newcommand{\pre}{\mathrm{pre}}
\newcommand{\post}{\mathrm{post}}

\theoremstyle{remark}
\newtheorem{definition}[theorem]{Definition}

\newtheorem*{remark}{Remark}

\begin{document}

\title{Euclidean mirrors and first-order changepoints in network time series}

\author[1]{Tianyi Chen}
\author[2]{Zachary Lubberts\thanks{Corresponding Author: zlubberts@virginia.edu}}
\author[1]{Avanti Athreya}
\author[3]{Youngser Park}
\author[1]{Carey E. Priebe}

\affil[1]{Department of Applied Math and Statistics, Johns Hopkins University}
\affil[2]{Department of Statistics, University of Virginia}
\affil[3]{Center for Imaging Science, Johns Hopkins University}

\date{} %

\maketitle

\begin{abstract}
We describe a model for a network time series whose evolution is governed by an underlying stochastic process, known as the latent position process, in which network evolution can be represented in Euclidean space by a curve, called the Euclidean mirror. We define the notion of a first-order changepoint for a time series of networks, and construct a family of latent position process networks with first-order changepoints. We prove that a spectral estimate of the associated Euclidean mirror localizes these changepoints, even when the graph distribution evolves continuously, but at a rate that changes. Simulated and real data examples on brain organoid networks show that this localization captures empirically significant shifts in network evolution.  

\vspace{1em}
\end{abstract}

\tableofcontents

\section{Introduction}\label{sec:intro}
Identifying structural changes for a time series of network data is a critical inference task in modern statistics and interdisciplinary data science. For example, 
in organizational networks, statistical changepoint detection can identify transformations induced by the COVID-19 pandemic in communication patterns of corporations \cite{zuzul2021dynamic}. In brain organoid networks, changepoint detection can distinguish the biologically significant emergence of inhibitory neurons and growth of astrocytes \cite{chen2023discovering}. As another illustration, spectral network analysis can discern marked fluctations in global commodity prices in a time series of agricultural trade networks \cite{wang2023multilayer}. In all of these cases, the data is a collection of time-indexed
networks, each comprised of nodes and edges. The analysis of such time series requires tractably modeling network evolution and managing the inherently non-Euclidean nature of network data.

This non-Euclidean structure, along with the dependency between edges within and across networks, as well as the high dimensionality and noise that accompany Euclidean representations of network features, all present challenges for the development of principled methodology for network time series. As a consequence, while changepoint analysis for Euclidean data has a rich history \cite{shumway2000time,box2015time,kedem2005regression, verzelen2023optimal, padilla2021optimal}, time series analysis for networks is comparatively new \cite{zhou2024survey}. 
A natural point of departure is the spectral analysis of a sequence of adjacency matrices or a network tensor: see
\cite{paul2020spectral,jing2021community,pantazis2022importance,macdonald2022latent,wang2023multilayer,athreya2025euclidean,agterberg2022joint,padilla2022change, lin2024dynamic} for approaches to the estimation of common parameters, hypothesis testing for differences in networks, and analyses of changepoints and anomalous behavior across networks.

To formalize these ideas, we consider a collection of random networks on the same vertex set, indexed by time. For this network time series, we assume that associated to each vertex $i$ in each network at time $t$ is a (typically unobserved) low-dimensional vector, called the {\em latent position}, and that the probability of connections between nodes in the network depends on these latent positions.
We assume these latent positions evolve according to some underlying stochastic process, called a {\em latent position process} (LPP). A central question is whether statistical analysis of observed network data can identify changes in the latent position process. 

This observed network data consists of a time-indexed collection of adjacency matrices, each entry of which is a Bernoulli random variable with success probability determined by the latent positions of the respective nodes.
The observed network adjacency $A_t$ at time $t$  is a noisy version of the matrix of connection probabilities $P_t$ at time $t$.
The matrix of connection probabilities $P_t$ is itself a compression of the information in the realizations of the latent position processes.
We do not observe the $P_t$ matrices, much less the realizations of the latent position process.
However, because the edge distribution of a latent position network is governed by the LPP, changepoints for the distribution of the network can be associated to changepoints in the latent position process. 
Specifically, we consider changes in the distribution of the process itself, which we call {\em zeroth} order changepoints, as well as changepoints in the distribution of the increments of the process, which we term {\em higher-order changepoints}. 
We argue that for real-world networks, evolution in the connection probabilities is the norm, rather than the exception, so the appropriate notion of a changepoint should not be restricted only to a change in the network's distribution, but include potentially higher-order changes in its pattern of evolution. Related changepoint detection for real-valued stochastic processes, such as drift terms in Brownian motion, is a classical problem; see, for instance, \cite{shiryaev1996minimax, shiryaev1963optimum, pollak1985diffusion, pollak2009optimality}. The data we observe is not the underlying latent position process itself, but is instead the realization of the networks generated by the latent position process; the relevant inference task is to detect appropriate changes in the latent position process based on the time-varying observations of these networks.

To localize higher-order changepoints, we follow the methodology in~\cite{athreya2025euclidean}, where the authors develop a model for time series of networks generated by an LPP and associate to this LPP a finite-dimensional Euclidean curve, called a {\em mirror}, which provides a Euclidean representation for the network evolution over time. In \cite{athreya2025euclidean}, spectral decompositions and classical multidimensional scaling (CMDS) on the observed time series of networks provide consistent estimates of this mirror. When the mirror exhibits manifold structure, nonlinear dimension reduction techniques, such as isometric mapping (ISOMAP) \cite{tenenbaum2000global}, yield approximations to this manifold, which we dub the \emph{iso-mirror}. The mirror and iso-mirror represent important features of the LPP and the dynamics of the time series of networks, and shifts in the mirror or iso-mirror can reflect corresponding changes in the latent position process. 

\begin{figure}[ht]
\centering
\includegraphics[width=1.9in]{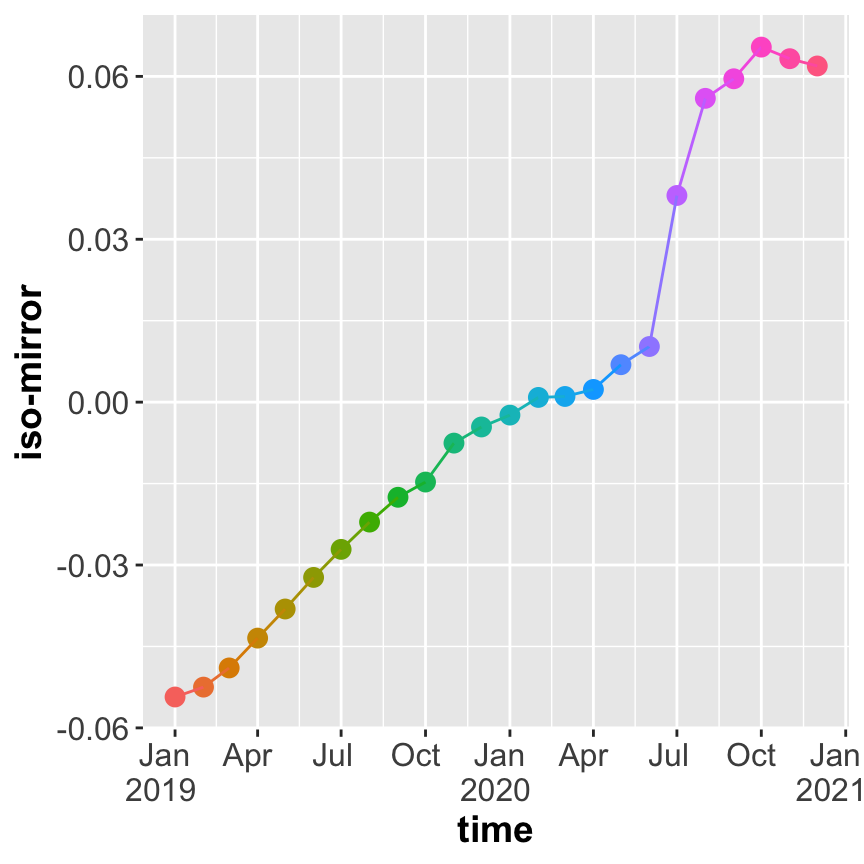}
\includegraphics[width=1.9in]{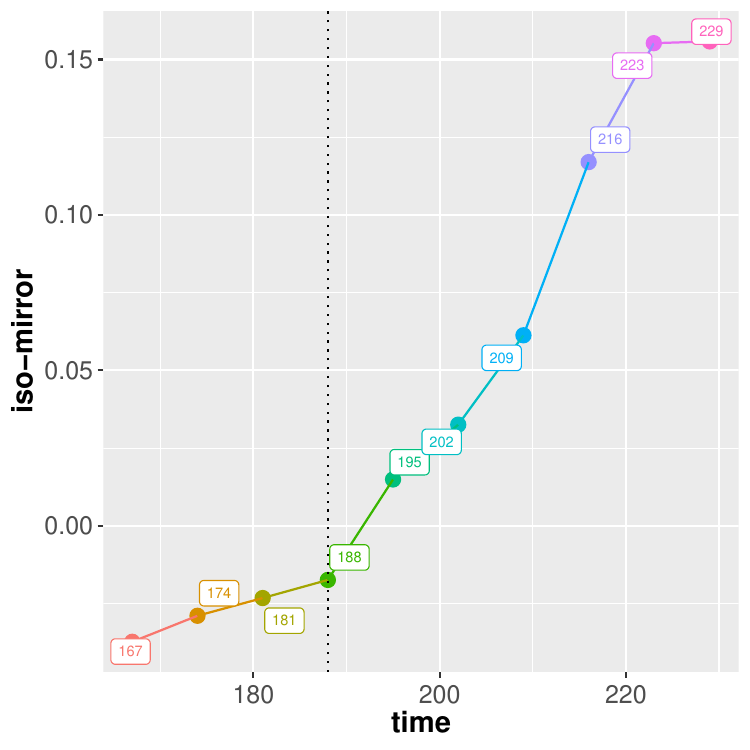}
\caption{Piecewise linearity in a real data example, reproduced from \cite{athreya2025euclidean}. Left panel shows the iso-mirror estimation for a time series of organizational communication networks generated each month from January 2019 to January 2021. A linear trend observed prior to April 2020, after which COVID-19 work-from-home protocols were introduced. Right panel shows iso-mirror estimation on time series of brain organoid connectivity networks sampled roughly once per week. Estimated iso-mirror exhibits piecewise linearity with slope change at day 188. 
}
\label{fig:real-data}
\end{figure}

Fig \ref{fig:real-data} showcases two real data examples from \cite{athreya2025euclidean} and \cite{chen2023discovering}, with plots of iso-mirrors over time. We observe that in both cases, the iso-mirrors exhibit approximate piecewise linearity. The left panel is the iso-mirror for a time series of communication networks at Microsoft during the initial months of pandemic-induced remote work. A notable linear trend is evident before a significant shift around June 2020 \cite{zuzul2021dynamic}. Similarly, the right panel shows the iso-mirror of inferred effective connectivity for brain organoid connectomes. It too displays a piecewise linear trend with a change in the slope at day 188 \cite{chen2023discovering}, which coincides with the development of inhibitory neurons and the growth of astrocytes.

Motivated in part by this piecewise linearity, we construct a latent position process network with an asymptotically piecewise linear mirror and a first-order changepoint. The LPP is based on a random walk and the first-order changepoint is introduced by shifting the jump probability at $t^*$ from $p$ to a different value $q \neq p$. We prove that the associated  mirror converges uniformly, as the number of networks in our times series increases, to a piecewise linear function with a slope change---namely, a higher-order changepoint---at $t^*$. Therefore, the task of identifying changepoints within network time series translates to analyzing slope changes in the mirror, allowing us to frame changepoint detection in networks as a changepoint analysis problem in Euclidean space. The asymptotic piecewise linearity property of the mirror leads to a localization estimator for $t^*$, and we prove that this estimator is consistent.
We demonstrate this localization procedure on simulated and real data, and show that networks can differ markedly before and after these higher-order changepoints. Higher-order changepoint localization thus enables the discovery of finer changes in network evolution.

\subsection{Comparison to previous work}

Latent position network models \cite{jones2020multilayer, wang2023multilayer, padilla2022change, agterberg2022joint, athreya2025euclidean, kei2024change} are a common point of departure for both individual and joint network analysis. These include the stochastic block model \cite{bhattacharjee2020change,padilla2022change,fan2022alma}, the generalized random dot product graph, and more intricate latent space models, such as multiplex and multilayer models \cite{macdonald2022latent,wang2023multilayer,jones2020multilayer,pensky2024clustering, noroozi2022sparse}. %
In network time series, in addition to a generative model for the network at each time point, it is necessary to describe the dependence between networks at different time points, and the approach to this varies. In \cite{macdonald2022latent,jones2020multilayer,wang2023multilayer, lin2024dynamic}, the authors propose shared structural framework for the latent positions of nodes across different layers and times. In \cite{macdonald2022latent}, the latent position matrix for each layer or time is split into initial columns representing common latent positions and subsequent columns for individual latent positions that differ across layers or times. In \cite{lin2024dynamic}, the authors use kernel debiased sum-of-squares spectral decompositions to estimate community memberships in stochastic blockmodels where block memberships of can shift randomly with time. In \cite{wang2023multilayer}, the authors consider a multilayer random dot product model where all layers or times share identical latent positions, but the connectivity matrices determining connection probabilities from these latent positions are allowed to vary. In \cite{padilla2022change}, the authors model inter-network dependence by setting the latent positions $X_i(t)$ for a node $i$ at time $t$ as
$$
X_i(t)\begin{cases}
=X_i(t - 1), & \text{with probability } p, \\
\sim F, & \text{with probability } 1 - p,
\end{cases}
$$
where $F$ is the distribution for the latent position at $t-1$. 
In \cite{kei2024change}, a multi-layer perceptron (MLP) is employed to produce a latent position matrix at each time $t$ from a normal distribution $\mathcal{N}(\mu_t,I_d)$, with MLP parameters shared across time. 
Further, several studies assume independence of networks across layers or times \cite{wang2021optimal,bhattacharjee2020change}. We take the view that common subspace assumptions may be natural in multilayer network settings, but are implausible in dynamic networks, especially over long time intervals. When each individual's latent position varies over time, common subspace assumptions typically will not hold.

To appropriately model changepoints in network time series, we must delineate how network dynamics evolve over time and investigate different types of changepoints; this motivates our definition of changepoints of different orders. As a baseline model (see  \cite{shiryaev1996minimax, shiryaev1963optimum, pollak2009optimality, pollak1985diffusion,wang2023multilayer,kei2024change}) consider what we define as a {\em zeroth-order} changepoint: 
\begin{model}[Zeroth-order Changepoint]
\label{model:mean-shift}
The time series of graphs $\{\mathbf{G}_t\}^{m}_{t=1}$ having marginal distributions $\{\mathcal{L}_t\}^{m}_{t=1}$ has a \emph{zeroth-order changepoint} at $t^*, 1<t^*<m$, when 
$$
\mathcal{L}_{t-1} \neq \mathcal{L}_t \text{~if and only if~} t=t^*. 
$$
\end{model}

For time series of networks, this is the kind of changepoint that has traditionally been investigated in the literature. In \cite{wang2021optimal}, the underlying probability matrices change only at the changepoints; the authors address detection and localization of an unknown number of changepoints and introduce two methods, both relying on cumulative sum (CUSUM) statistics that necessitate dividing the time series data into two separate series.  The authors show that under certain conditions, these methods achieve minimax optimality. Similarly, \cite{padilla2022change} concerns a model where the latent position distribution again changes solely at the changepoints, and the authors deploy spectral methods and CUSUM statistics for localization. 
In \cite{wang2023multilayer}, changepoints in multilayer random dot product models are zeroth order, namely time points where the connectivity matrices change. The authors develop techniques for online changepoint detection using scan statistics derived from tensor decompositions of the adjacency matrices in the network time series. In \cite{kei2024change}, a changepoint is defined as a shift in the expected value $\mu_t$ of the distribution generating the latent position matrix. The authors determine $\hat{\mu}_t$ using maximum approximate likelihood estimation, incorporating group-fused lasso regularization, and show that careful analysis of the sequential differences $\hat{\mu}_t-\hat{\mu}_{t-1}$ can effectively detect and localize changepoints.
We note that outside of the statistical network analysis literature, there has been significant methodological development of changepoint detection in dynamic networks, as described in \cite{zhou2024survey} and references within.

In the present work, we consider a time series of networks in which each network is a random dot product graph on the same vertex set with latent positions belonging to a low-dimensional Euclidean space of fixed dimension.
We assume the latent positions are samples from a latent position process, inducing dependence across time. A distinctive aspect of our model is that it permits the underlying distribution for networks to drift, with the {\em rate} of drift altering at the changepoint; this is what we formally define as a first-order changepoint in Section \ref{Sec:main model and prior results}. This model renders the algorithms from aforementioned work less effective in localizing our changepoint. Nevertheless, our Euclidean mirror methodology captures essential signal in the latent position process, leading to consistent changepoint localization.

Finally, we note that the present work builds on the Euclidean mirror framework and latent position process time series of graphs model introduced in \cite{athreya2025euclidean}. The novelty of the present work consists of the following: we define a notion of higher-order changepoints for network time series, and find a tractable latent position process with a first-order changepoint for which we can analytically derive the Euclidean mirror. We prove first-order changepoint localization results for not just this particular model, but a class of latent position processes that exhibit piecewise-linear structure in their Euclidean mirrors, which we have observed appear in multiple application domains. Moreover, while \cite{athreya2025euclidean} only considers the case where the number of time-points is fixed, in the present work we study the case where the number of time-points goes to infinity while remaining within a bounded interval, the so-called ``in-fill asymptotics." Together, these results significantly improve our theoretical understanding of inference for network time series.

\subsection{Organization of paper}

We organize the paper as follows. In Section \ref{Sec:main model and prior results}, we first introduce our model for network time series; second, we construct a latent position process with a first-order changepoint based on a random walk with a shift in jump probability at a certain time point $t^*$; and third, we prove that its associated mirror exhibits an asymptotic piecewise linear structure, with a slope change at $t^*$. 
In Section \ref{Sec:changepoint analysis} we use spectral techniques and classical multidimensional scaling to estimate the mirror based on network observations, and we exploit the asymptotic piecewise linearity of the underlying mirror to construct an $l_\infty$ localization estimator for the changepoint $t^*$. We prove that when asymptotic piecewise linearity holds, this estimator is consistent. In numerical experiments, we demonstrate that the estimated iso-mirror can recover the changepoint, even with a limited sample size $n$. We also investigate the bias-variance tradeoff in changepoint estimation arising from the choice of embedding dimension for the estimated mirror, and discuss the changepoint detection problem, demonstrating that our estimated mirror may also be useful for this task. Finally, we examine a real data time-series of brain organoid networks, observing a piecewise linear structure for mirror estimates in this data as well, and localizing a changepoint. In Section \ref{Sec:Con_and_dis}, we discuss ongoing research and further work. All proofs are in the Appendix.

\section{Main model and prior results}
\label{Sec:main model and prior results}

\subsection{Notation}
Consider a vector $v \in \R^d$, which is represented as a column vector. We express $v$ as $v=\left(v_1, v_2, \cdots, v_d\right)^\top = [v_i]_{i=1}^d$, where $v_i$ denotes the $i$th entry of $v$ or sometimes we also use $v(i)$ to denote the $i$th entry of $v$. The vector Euclidean norm is denoted by $\|\cdot\|$. For a matrix $A \in \R^{p_1 \times p_2}$, the element at the $i,j$th position is represented as $A_{i,j}$, and $A$ can also be denoted by $[a_{ij}]_{i=1,j=1}^{p_1,p_2}$ with $a_{ij}$ as the $i,j$th entry.
The spectral norm of $A$ is denoted by $\|A\|_2$, the Frobenius norm by $\|A\|_F$, the maximum Euclidean row norm by $\|A\|_{2 \to \infty}$, and the maximum absolute row sum by $\|A\|_{\infty}$. For a square matrix $A \in \R^{n \times n}$ with real eigenvalues, $\lambda_1(A)$ denotes the most positive eigenvalue, $\lambda_2(A)$ the second most positive eigenvalue, and so on. We use $\Omega$ to denote a sample space, $\mathcal{F}$ a $\sigma$-algebra of events, and $\mathbb{P}$ a probability measure.

We require the following definitions on asymptotic order and convergence.

\begin{definition}[Order notation]
When $\omega(m)$, $\alpha(m)$ are two quantities depending on $m$, we say that $\omega$ is of order $\alpha(m)$ and use the notation $\omega(m) \sim \Theta(\alpha(m))$ if there exist positive constants such that for $m$ sufficiently large, $c\alpha(m)\leq \omega(m) \leq C \alpha(m)$. We write $\omega(m) \sim O(\alpha(m))$ if there exists a constant $C$ such that for sufficiently large $m$, $\omega(m) \leq C \alpha(m)$.
\end{definition}

\begin{definition}[Convergence with high probability]
Given a sequence of events $\left\{F_n\right\} \in \mathcal{F}$, where $n=1,2, \cdots$, we say that $F_n$ occurs with high probability, and write $F_n$ w.h.p., if for some $r>1$, there exists a finite positive constant $C$ depending on $r$ such that $\mathbb{P}\left[F_n^c\right] \leq C n^{-r}$ for all $n$. 
\end{definition}

\subsection{Model formulation and key definitions}\label{Sec:Model formulation and key definitions}
In our model for network evolution, we build on the framework for network time series in \cite{athreya2025euclidean} and consider generalized random dot product graphs (GRDPG), which are sufficiently flexible to approximate most independent-edge networks (see \cite{tang2012universally} and \cite{wolfe13:_nonpar}).

\begin{definition}[Random dot product graph]
We say that the undirected random graph $G$ with adjacency matrix $\bA\in\RR^{n\times n}$ is a \emph{random dot product graph (RDPG)} with latent position matrix $\bX\in\RR^{n\times d}$, whose rows are the vectors $X^1,\ldots,X^n\in\mathcal{X}\subseteq\RR^d$, if
$$\PP[\bA|\bX]=\prod_{i<j}\langle X^i,X^j \rangle^{A_{i,j}}(1- \langle X^i,X^j \rangle)^{1-A_{i,j}},$$
where $\langle x,y\rangle= x^\top y$ is the inner product between two column vectors. We call $\bP=\bX \bX^T$ the connection probability matrix. 
If instead $\bP = \bX I_{p,q}\bX^\top$, where $I_{p,q}=I_p\oplus(-I_q)$ for some $p+q=d$, we call $\bA$ a \emph{generalized random dot product graph (GRDPG).}
\end{definition}

\begin{remark}[Orthogonal nonidentifiability in RDPGs] \label{rem:nonid}
Note that if $\bX \in \R^{n \times d}$ is a latent position matrix
and $\bW \in \R^{d \times d}$ is orthogonal,
$\bX$ and $\bX\bW$ give rise to the same distribution over graphs.
Thus, the RDPG model has a nonidentifiability up to orthogonal transformation. In the case of a GRDPG, $\bX$ and $\bX\bW$ give rise to the same distribution whenever $\bW I_{p,q} \bW^\top=I_{p,q},$ meaning that $\bW$ is an indefinite orthogonal transformation.
\end{remark}

To model randomness in the underlying features of each vertex, we consider latent positions that are themselves random variables defined on a probability space $(\Omega, \mathcal{F}, \PP)$. The notion of an inner product distribution ensures well-defined probabilities.

\begin{definition}[Inner product distribution]
\label{def:innerprod}
Let $F$ be a probability distribution on $\R^d$. we call $F$ a \emph{$d$-dimensional inner product distribution}
if $0 \leq x^{\top} y \leq 1$ for all $x,y \in \supp F$. We call $F$ a \emph{$(p,q)$-dimensional generalized inner product distribution} if $0\leq x^\top I_{p,q} y \leq 1$ for all $x,y\in\supp F$, where $I_{p,q}=I_p\oplus (-I_q)$. 
\end{definition}

Suppose that the time series of networks come from an RDPG model, where at each time $t$, the latent positions of each vertex are drawn independently from a common distribution $F_t$. To describe such a family of networks indexed by 
time, we consider a {\em latent position stochastic process} and its corresponding time series of networks.

\begin{definition}[Latent position process]\label{def:latent_pos_proc}
A {\em latent position process} $\varphi(t)$ is a map $\varphi:[0,T]\rightarrow L^2(\Omega, \mathbb{P})$ such that for each $t\in [0,T]$, $\varphi(t)=X_t$, a random vector in $\R^d$ which has a (generalized) inner product distribution.
\end{definition}

\begin{definition}[Latent position process network time series]
\label{def:tsg}
Let $\varphi$ be a latent position process, and fix a given number of vertices $n$ and collection of times $\mathcal{T}\subseteq[0,T]$. We draw an i.i.d.\ sample $\omega_j\in \Omega$ for $1\leq j\leq n$, and obtain the latent position matrices $\bX_{t}\in\RR^{n\times d}$ for $t\in\mathcal{T}$ by appending the rows $X_t(\omega_j)$, $1\leq j\leq n$. The \emph{time series of graphs (TSG)} $\{G_t: t\in\mathcal{T}\}$ are conditionally independent RDPGs with latent position matrices $\bX_{t}, t\in\mathcal{T}$.
\end{definition}

\begin{figure}
    \centering
    \begin{tikzpicture}[decoration=snake]
	\draw[scale=2,cm={cos(-45),sin(-45),-sin(-45),cos(45),(0cm,0cm)}] (0,0) arc[thick,start angle=270,end angle=315,x radius=3,y radius = 6] node foreach \p in {0,0.1,...,1.1} [pos=\p,inner sep=0.5pt,circle,fill=black]{} node (a) [inner sep=0.5pt,draw,fill=white,circle,pos=0.6]{\small $X_t$} node [pos=1,above right] {\small $L^2(\Omega)$} node [pos=0.3,above,rotate=-20]{\small $t\rightarrow$};
	\draw[thick,scale=2.1,cm={cos(-40),sin(-40),-sin(-40),cos(40),(0cm,-0.75cm)}] (0,0) arc[start angle=270,end angle=320,x radius=3,y radius=6] node foreach \p in {0,0.1,...,1.1} [pos=\p,inner sep=1.5pt,rectangle,fill=black]{} node (b) [inner sep=2pt,rectangle,draw,fill=white,pos=0.6]{\small $\bX_t$} node [pos=1,above right] {\small $\RR^{n\times d}$};
	\draw[very thick, dashed, scale=2.25,cm={cos(-35),sin(-35),-sin(-35),cos(35),(0cm,-1.5cm)}] (0,0) arc[start angle=270,end angle=325,x radius=3,y radius=6] node foreach \p in {0,0.1,...,1.1} [pos=\p,inner sep=2pt,rectangle,fill=black]{} node (c) [inner sep=2pt,rectangle,draw,solid,fill=white,pos=0.6]{\small $\mathbf{A}_t$} node[pos=1,above right] {\small $\{0,1\}^{n\times n}$};
	\draw[->] (a) -- (b) node[pos=0.3,right]{IID Sample};
    \path (a) node[above=16pt] {$\{X_t:t\in[0,T]\}$};
	\draw[->,decorate] (b) -- (c) node[pos=0.3,right]{Observe graph};
    \end{tikzpicture}
    \caption{A schematic view of a latent position process time series of graphs. Trajectories follow a latent position process, so marginal distributions of $X_t$ are dependent between times. Each IID sample from the LPP provides a row to each matrix $\bX_t$ across time. Conditioned on the latent position matrices, the adjacency matrices are observed conditionally independently for each time. Solid lines denote dependence, while dotted lines represent (conditional) independence.}
    \label{fig:lpptsgschematic}
\end{figure}

 Existing literature on changepoint detection for networks has primarily focused on the marginal distributions of the networks. On the other hand, for a TSG where edges are assumed to arise conditionally independently (at each time and between times), and with a fixed small rank $d$ (or $p,q$ with $p+q=d$) for the mean matrices at each time, there is implicitly a (generalized) latent position process governing network evolution. As such, it is natural to define changepoints through properties of the LPP, rather than the distributions of the graphs themselves. This leads us to the notion of changepoints of different orders, below.

\begin{definition}
\label{def:highchangepoints}
A latent position process $\varphi(t)$ is said to have a \emph{zeroth-order changepoint} if there is some $t^*\in[0,T]$ such that
$$ X_t - X_{t'} \overset{\mathcal{L}}{=} \begin{cases}\delta_0 &\text{if }t,t'\leq t^*\text{ or }t^*\leq t,t'\\ F \neq \delta_0&\text{otherwise.}\end{cases}$$
Here $\delta_0$ is point mass at $0\in\RR^d$, and $F$ is some other distribution on $\RR^d$.

Given $\delta>0$, define the $k$th order difference of $\varphi$ at mesh $\delta>0$ recursively as follows:
\begin{align*}
\Delta_1^\delta(t) &= X_t - X_{t-\delta},& t&\in[\delta,T],\\ 
\Delta_k^\delta(t) &= \Delta_{k-1}^\delta(t)-\Delta_{k-1}^\delta(t-\delta),& t&\in[k\delta,T].
\end{align*}

A latent position process $\varphi(t)$ is said to have a \emph{$k$th-order changepoint of mesh }$\delta$ if there is some $t^*\in[0,T]$ such that for some distributions $F_1$, $F_2$, $F_3(t)$, $t^{*}<t\leq t^{*}+\delta$, not all equal, we have 
$$ \Delta_{k}^\delta(t) \overset{\mathcal{L}}{=} \begin{cases} F_1 &\text{if }t\leq t^*,\\
F_2&\text{if }t^*+\delta<  t,\\ F_3(t)&\text{if }t^*< t\leq t^*+\delta.\end{cases}$$ 
\end{definition}

Note that a first-order changepoint exactly corresponds to the setting in which 
$$\{\varphi(t): t\in [0,t^*)\}, \{\varphi(t): t\in(t^*+\delta,T]\}$$ are stationary stochastic processes (at least with respect to increments of length $\delta$), but with a loss of stationarity at the point $t^*$. Our definition of a higher-order changepoint is analogous to a finite-differences
approximation to a slope and higher-order derivatives of a real-valued function.

We emphasize in Definition \ref{def:tsg} that each vertex in the TSG corresponds to a single $\omega\in\Omega$, which induces dependence between the latent positions for that vertex across times, but the latent position trajectories of any two distinct vertices are independent of one another across all times. Since these trajectories form an i.i.d.\ sample from the latent position process, it is natural to measure their evolution over time using the metric on the corresponding random variables, as described in \cite{athreya2025euclidean}. We reproduce the maximum directional variation metric, $d_{MV}$, a metric on the space of square-integrable random variables, as follows.
\begin{definition}
The \emph{maximum directional variation metric} $d_{MV}:L^2(\RR^d)^2\rightarrow [0,+\infty)$ is given by
\begin{equation}
d_{MV}(X_t,X_{t'}):=\min_{W \in \mathcal{O}^{d \times d}} \left\|\EE[(X_t-WX_{t'})(X_t-WX_{t'})^\top]\right\|_2^{1/2},
\end{equation}
where $\mathcal{O}^{d\times d}$ is the set of all $d \times d$ orthogonal matrices and $\| \cdot \|_2$ is the spectral norm.
\end{definition}

In the definition of this distance, the expectation is over $\omega\in\Omega$ determining the trajectory of $X_t$ over time, so this distance depends on the joint distribution of $X_t$ and $X_{t'}$. In particular, $d_{MV}$ depends on more than just the marginal distributions of the random vectors $X_t$ and $X_{t'}$ individually; it also takes into account the dependence inherited from the latent position process $\varphi$. 

Next we consider the geometric properties of the image $\varphi([0,T])$, equipped with the metric $d_{MV}$. Often, the map $\varphi$ admits a Euclidean analogue, called a {\em mirror}, which is a finite-dimensional curve that retains important signal from the generating LPP for the network time series. 

\begin{definition}[Approximate Euclidean realizability with mirror $\psi$]
Let $\varphi$ be a latent position process. For a fixed $\alpha \in (0, 1)$, we say that $\varphi$ is \emph{approximately \(\alpha\)-H\"older Euclidean \(c\)-realizable} if $\psi$ is \(\alpha\)-H\"older continuous, and there is some $C > 0$ such that
\[
\biggl| d_{MV}(\varphi(t), \varphi(t')) - \|\psi(t) - \psi(t')\| \biggr| \leq C|t - t'|^\alpha \quad \text{for all } t, t' \in [0, T].
\]
When $C=0$, we say that $\varphi$ is \emph{exactly Euclidean $c$-realizable.}
\end{definition}

\begin{remark}
Euclidean realizability implies the pairwise $d_{MV}$ distances between the latent position process at $t$ and $t'$ can be approximated with Euclidean distances along the curve $\psi$ at $t$ and $t'$. This curve reflects, in Euclidean space, the $d_{MV}$ distances, and hence we call $\psi$ the {\em Euclidean mirror}.
\end{remark}

A simple special case is where all latent positions follow a single deterministic trajectory:

\begin{lemma}\label{lem:deter_lemma}
If the latent position process $\{\varphi(t)\}$ is deterministic and $t \in (a,b)$, $\varphi(t) \in \R^d$, then it is exactly Euclidean 1-realizable, with mirror $\psi(t)=\|\varphi(t)\|_2-\frac{\int^b_a \|\varphi(s)\|_2\, ds}{b-a} $. 
\end{lemma}

When the LPP is exactly $c$-Euclidean realizable with mirror $\psi$, and we sample $m$ time points
$$\mathcal{T}=\{t_1,t_2,\cdots,t_m\} \subseteq [0,T],$$ the corresponding distance (or dissimilarity) matrix $$\left(\mathcal{D}_\varphi\right)_{i,j}:=d_{MV}(X_{t_i},X_{t_j})$$ is an \emph{exactly $c$-Euclidean realizable distance matrix}: that is, there exist $m$ points $$\psi(t_1),\psi(t_2),\cdots,\psi(t_m) \in \mathbb{R}^c$$ such that $$\left(\mathcal{D}_\varphi\right)_{i,j}=\|\psi(t_i)-\psi(t_j)\| \text{~for all~} i,j \in \{1,2,\cdots,m\}.$$ 
In this case, applying classical multidimensional scaling (CMDS) to $\mathcal{D}_{\varphi}$ will recover the mirror $\psi(t_1),\psi(t_2),\cdots,\psi(t_m)$ exactly up to an orthogonal transformation. Recall that the CMDS embedding is defined as follows.

\begin{definition}[CMDS embedding to dimension $c$]
\label{def:CMDS}
Let $\mathcal{D} \in \mathbb{R}^{m \times m}$ be a distance matrix and define the centering matrix $P$ by $P:=I_m-\frac{J_m}{m}$ where $I_m$ is the $m \times m$ identity matrix and $J_m$ is the $m \times m$ all ones matrix. Consider $B:=-\frac{1}{2}P\mathcal{D}^{(2)}P$ where $\mathcal{D} ^{(2)}$ is the entrywise square of the distance matrix $\mathcal{D} $. Denote the $c$ most positive eigenvalues and corresponding orthogonal eigenvectors of $B$ as $\lambda_1,\cdots,\lambda_c$ and $u_1,\cdots,u_c$. Set $\max(\lambda_i,0)=\sigma^2_i$ and $S$ a diagonal matrix with $S_{ii}:=\sigma^2_i$ for $i \in \{1,\dots,c\}$; let $U \in \R^{m \times c} := [ u_1 | \cdots | u_c ]$. The classical multidimensional scaling of $\mathcal{D}$ into dimension $c$, where $1\leq c\leq m-1$, is the set of $m$ points $\psi(t_1),...,\psi(t_m) \in \mathbb{R}^c$ defined by the rows of $\Psi$ as 
$$
\Psi:=[\psi_{j}(t_i)]_{i=1,j=1}^{m,c}
=\left[
\begin{array}{ccc}
\psi_1(t_1) & \cdots & \psi_c(t_1) \\ 
\vdots &  & \vdots \\ 
\psi_1(t_m) & \cdots & \psi_c(t_m)
\end{array}\right]=
 \left[
 \begin{array}{c}
  \psi^{\top}(t_1) \\ 
 \vdots \\ 
 \psi^{\top}(t_m) 
 \end{array}
 \right].
$$
We will also write $\Psi=
US^{\frac{1}{2}}=
\left[ \sigma_1 u_1|\cdots| \sigma_c u_c\right]$ to refer to the columns of this matrix.
\end{definition}

\begin{remark}
If there are no positive eigenvalues for \( B \), then \( S=0 \) and thus \( \Psi=0 \). This implies that the distance matrix has no meaningful Euclidean representation. As highlighted in Theorem \ref{thm:PSD-iffrealizable}, \( B \) must be positive semidefinite for \( \mathcal{D} \) to be exactly Euclidean realizable.
\end{remark}
The next theorem from \cite{trosset2020rehabilitating} gives necessary and sufficient conditions for a distance matrix to be exactly Euclidean realizable.
\begin{theorem}\label{thm:PSD-iffrealizable}
A distance matrix $\mathcal{D}$ is exactly $c$ Euclidean realizable, and is not $c-1$ Euclidean realizable, if and only if $B=-\frac{1}{2}P\mathcal{D}^{(2)}P$ is positive semidefinite with exactly $c$ positive eigenvalues.
\end{theorem}
Embedding a distance or dissimilarity matrix through CMDS provides a constellation of points in Euclidean space whose interpoint Euclidean distances approximate those in the dissimilarity matrix. In the case of a latent position process, this embedding can provide key signal about the underlying process itself. We call this the {\em zero-skeleton} mirror.
\begin{definition}[Zero-skeleton mirror]\label{def:0-skeleton-dMV-mirror} Suppose $\mathcal{D}_{\varphi}$ is the $d_{MV}$ distance matrix associated with an  LPP $\varphi$. The output of CMDS applied to $\mathcal{D}_{\varphi}$ produces $m$ points in $\R^c$, denoted $\{\psi(t_1),...,\psi(t_m)\}$, called the \emph{$c$-dimensional zero-skeleton mirror}. 
\end{definition}

As we have seen in the analysis of TSGs from multiple application domains, the mirror is often approximately 1-d Euclidean realizable and exhibits piecewise linear structure. In the following section, we  introduce a class of LPP models which give rise to an analytically computable, asymptotically piecewise linear mirror with a first-order changepoint, and we use spectral decompositions of the observed networks to localize this changepoint. The formal construction of a specific LPP with such piecewise linear structure allows us to develop principled methodology for changepoint localization in these settings.

\subsection{Random walk latent position processes}
\subsubsection{A random walk process without changepoints}

In this section, we introduce a simple class of LPPs with a closed-form $d_{MV}$ distance. This enables us to analytically compute the mirror obtained through CMDS. 

\begin{model}[Random walk latent position process]\label{no-changepoint}
Let $m$ be an integer, $m\geq 2$. Let $t$ be an integer with $0 \leq t \leq m$. Let $c \geq 0, \delta_m>0$ be two constants satisfying $c+\delta_m m \leq 1$. For a fixed ``jump probability" $p \in (0,1)$, define a LPP $\varphi_m$ as follows:
\begin{align}
X_0 &=c \quad \text{with probability 1}, \notag \\
\text{For }t \geq 1, \quad X_t &=\begin{cases}
X_{t-1}+\delta_m &\quad\text{with probability }p, \\
X_{t-1} &\quad\text{with probability }1-p \notag.
\end{cases}    
\end{align}

\end{model}

We can, of course, express this latent position process as a familiar random walk: let $Z_i$, $i \geq 1$, be i.i.d random variables defined by $Z_i=\delta_m$ with probability $p$ and $Z_i=0$ with probability $1-p$; let $Z_0=c$ with probability one. Then for $t \in \{0,1,2,...,m\} $, $X_t= \sum^t_{k=0}Z_k$. Observe that for $t \geq 1$, $X_t-c=V_t\delta_m$, where $V_t$ is a Binomial random variable with $t$ trials and success probability $p$. Note that $X_t$ depends on $m$; in what follows, we suppress this dependence for notational convenience, but this will be important in subsequent sections when we let $m\rightarrow\infty$.

Since $c+\delta_m m \leq 1$, as $m$ increases, $\delta_m$ must decrease. For the rest of this section, we consider $X_0=0$ and $\delta_m=\frac{1}{m}$ so that our latent position process is supported on the set $\{0,\frac{1}{m}, \cdots, 1\}$. (When we consider a time series of graphs, we generally require $c>0$ and  $c+\delta_m m \leq 1$ to ensure sufficient signal strength in each network, but we adopt the specific choice of $c=0$ and $\delta_m=1/m$ in the present section for notational simplicity.)

Marginally at time $t$, the random dot product graph (RDPG) whose vertices have latent positions distributed according to $X_{t}$ is an $(m+1)$-block stochastic block model (SBM) with block assignment vector $\pi_{t,m,p}$ and block connection probability matrix $B_{m}$ defined by
$$
\pi_{t,m,p}(i)=
\begin{cases}
{\binom{t}{i}}p^i(1-p)^{t-i} & \text{for} \quad i \in \{0,1,2,...,t\} \\
0 & \text{for} \quad i \in \{t+1,t+2,...,m\},
\end{cases}
$$

$$
\left(B_{m}\right)_{i,j}=\frac{i j}{m^2} \quad \text{where} \quad i,j \in \{0,1,2,...,m\}.
$$
Note $B_{m}$ is a rank one matrix. The $d_{MV}$ distance is analytically tractable for Model \ref{no-changepoint}, and we compute it here. 

\begin{lemma}
\label{lem:dmv-nochangepoint}
For the latent position process in Model \ref{no-changepoint},
\begin{equation}\label{eq:dMV_Modelnochangepoint}
d^2_{MV}(X_t,X_{t'})=\left(\frac{p\left(t-t'\right)}{m}\right)^2+\left(p-p^2\right)\frac{|t-t'|}{m^2}\quad \text{for all } t, t' \in \{1,2,\cdots,m\},
\end{equation}
and this LPP is approximately \(\frac{1}{2}\)-H\"older Euclidean \(1\)-realizable with mirror $\psi(t)=\frac{pt}{m}$:
\begin{equation}\label{eq:approx_holder_Modelnochangepoint}
\biggr|d_{MV}(X_t,X_{t'})-\|\psi(t)-\psi(t')\| \biggr| \leq \sqrt{p-p^2}\frac{|t-t'|^{\frac{1}{2}}}{m} \quad \text{for all } t, t' \in \{1,2,\cdots,m\}.
\end{equation}

\end{lemma}

\begin{figure}[t]
    \centering
    \includegraphics[width=4.04in]{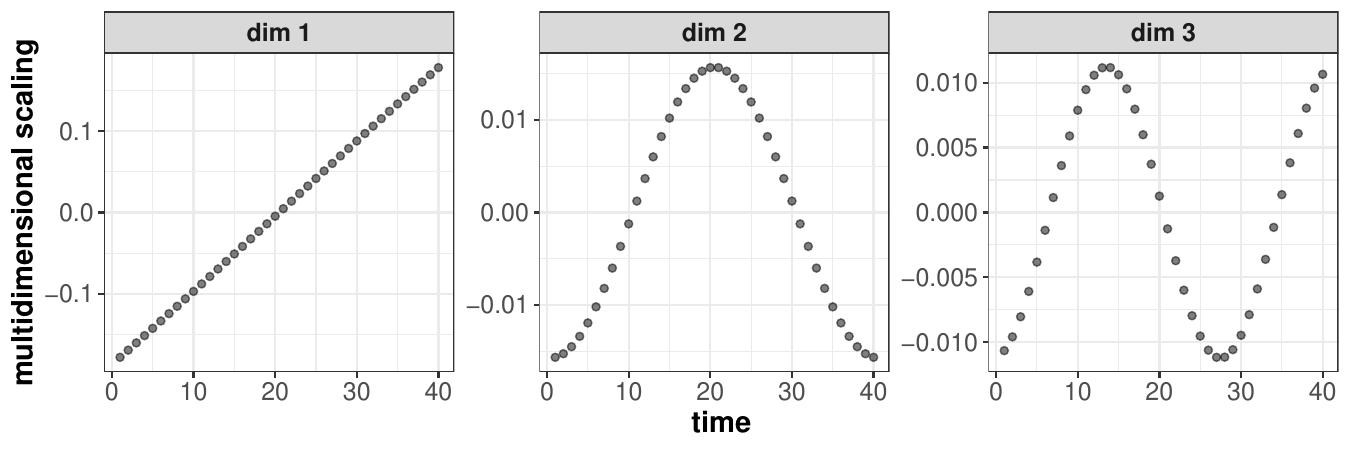}
    \caption{The first three dimensions of CMDS for the true distance matrix $\mathcal{D}_{\varphi_m}$ for Model \ref{no-changepoint}, where $c=0.1$, $\delta=\frac{0.9}{m}$, and $m=40$, $p=0.4$. 
    The left panel, the first dimension of CMDS, shows an approximately linear function of time increments $m$. The second and third dimensions, by contrast, are sinusoidal, but range over an order-of-magnitude smaller scaling. We provide analytic formulas for these in Appendix \ref{Sec:beyond-1st-dim}.}  
    \label{fig:cmds}
\end{figure}

In Figure \ref{fig:cmds}, we show the CMDS results for a distance matrix following Equation~(\ref{eq:dMV_Modelnochangepoint}) with model parameters $c=0.1$, $\delta=\frac{0.9}{m}$, $m=40$, and $p=0.4$. Clearly the first dimension of the mirror is close to the linear function $\psi(t)=p(\frac{t}{m}-\frac{1}{2})$. The second and third dimensions, by contrast, are sinusoidal, but range over an order-of-magnitude smaller scaling: these capture the absolute value term in Equation~(\ref{eq:dMV_Modelnochangepoint}).

\subsubsection{A random walk latent position process with first-order changepoint}\label{Sec:drift_firstorder_changepoint}

Having found the first dimension of the mirror for the LPP in Model \ref{no-changepoint}, we consider adding a changepoint to the LPP at time $t^*$ and examining whether this is effectively captured by the mirror itself, either in terms of localization of $t^*$ or detection of an underlying distributional change.
We introduce a more general LPP $\varphi_m$ which includes Model \ref{no-changepoint} as a submodel, but allows for a first-order changepoint via a modification of the jump probability at $t^*$.

\begin{model}[Random walk LPP with first-order changepoint]\label{changepointmodel} Let $m\geq 2$ be an integer. Let $c\geq0, \delta_m>0$ be constants satisfying $c+\delta_m m\leq 1$. We define the {\em random walk latent position process with first-order jump probability changepoint at $t_m^*$} by
\begin{align*}
X_0 &=c \quad \text{with probability 1,} \notag \\
X_t&=\begin{cases}
X_{t-1}+\delta_m \quad \text{with probability $p_t$} \\
X_{t-1} \quad \text{with probability $1-p_t$},
\end{cases}
\\
\text{where~} \text{$p_t = p$ for $t 
 \leq t_m^*$} & \text{ and $p_t = q$ for $t > t_m^*$}, \textrm{ where } p \neq q  \text{ and }  1\leq t,t_m^*\leq m.
\end{align*}
\end{model}

Set $c=0$ and $\delta_m=\frac{1}{m}$; we calculate the $d_{MV}$ distance in this case. Our next lemma exhibits the true $d_{MV}$ distance matrix for this process.

\begin{lemma}
\label{lem:dmv-changepoint}
For the latent position process in Model \ref{changepointmodel},\\
\noindent \adjustbox{width=4.75in}{$\displaystyle \left(\mathcal{D}^{(2)}_{\varphi_m}\right)_{i,j}=d^2_{MV}(X_i,X_j)=
\begin{cases}
p^2\left(\frac{i}{m}-\frac{j}{m}\right)^2+\frac{p-p^2}{m}|\frac{i}{m}-\frac{j}{m}| \quad i,j<t_m^*, \\
\left( p ( \frac{t_m^*}{m}- \frac{i}{m})+q(\frac{j}{m}-\frac{t_m^*}{m}) \right)^2 +\frac{p-p^2}{m} | \frac{t_m^*}{m}- \frac{i}{m}|+\frac{q-q^2}{m}|\frac{j}{m}-\frac{t_m^*}{m}| \quad i<t_m^*<j, \\
q^2\left(\frac{i}{m}-\frac{j}{m}\right)^2+\frac{q-q^2}{m}|\frac{i}{m}-\frac{j}{m}| \quad t_m^*<i,j. 
\end{cases}$}\\
This LPP is
approximately \(\frac{1}{2}\)-H\"older Euclidean \(1\)-realizable with mirror $$
\psi(t)=
\begin{cases}
    \frac{pt}{m} \text{~for~} t \leq t_m^*,\\
    \frac{qt}{m}+\frac{(p-q)t_m^*}{m} \text{~for~} t > t_m^*, \\
\end{cases} 
$$
and \begin{equation}\label{eq:approx_holder_Model_changepoint}
\biggr|d_{MV}(X_t,X_{t'})-\|\psi(t)-\psi(t')\| \biggr| \leq \frac{|t-t'|^{\frac{1}{2}}}{2m} \quad \text{for all } t, t' \in \{1,2,\cdots,m\}.
\end{equation}
\end{lemma}

\begin{figure}
    \centering
    \includegraphics[width=4.04in]{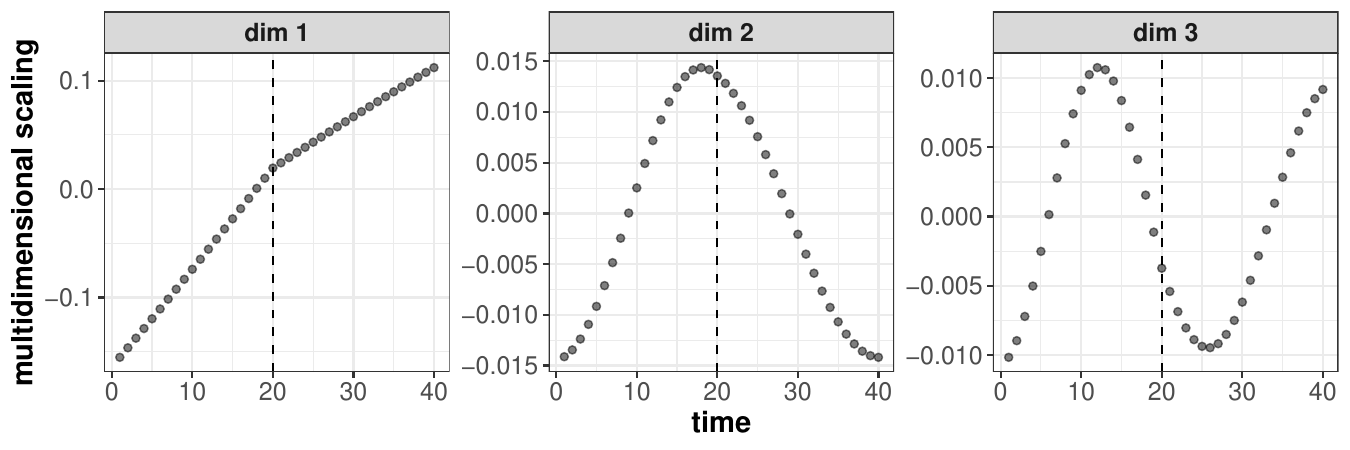}
    \caption{The first three dimensions of CMDS for the true distance matrix $\mathcal{D}_{\varphi_m}$ for Model \ref{changepointmodel}, where $c=0.1$, $\delta=\frac{0.9}{m}$, and $m=40$, $p=0.4$, $q=0.2$, and $t^{*}_m=20$. The first dimension, given in the leftmost panel, shows approximate piecewise linearity with a slope change at $t^{*}_m$. The second and third dimensions also exhibit a change in behavior at $t_m^*$, where the frequency of the cosine and sine curves changes abruptly, but the range in both sinusoidal terms is an order of magnitude smaller; analytical descriptions for these curves are provided in Appendix \ref{Sec:beyond-1st-dim}.
    }  
    \label{fig:cmds_changepoint}
\end{figure}

The proof of this lemma is similar to Lemma \ref{lem:dmv-nochangepoint}. As in the case without a changepoint, the absolute value terms in this function are of order $1/m$ as $m$ grows. This leads to a distance that is well-approximated by Euclidean distance along a one-dimensional piecewise linear curve with a change of slope at $t_m^*$. The results of CMDS applied to a distance matrix coming from this model are shown in Figure~\ref{fig:cmds_changepoint}. We devote the next section to the asymptotic behavior of these two models.

\subsubsection{In-fill asymptotics for random walk latent position processes}\label{S:asymppl}

Consider the asymptotics of estimation for latent position processes and associated mirrors with an increasing number of time points $t_1, \cdots, t_m$, all constrained to a fixed, bounded interval.
We now define the notion of \emph{asymptotically Euclidean 1-realizable} LPPs.
\begin{definition}[In-fill asymptotically Euclidean 1-realizable]\label{def:asy-1d}
Consider a latent position process $X_t$, sampled at $0 \leq t_1<\cdots< t_m \leq 1$ and denoted by $\mathcal{X}=\{X_{t_1}^{(m)},\ldots,X_{t_m}^{(m)}\}$, that is exactly Euclidean $(m-1)$-realizable with 0-skeleton mirror $\{\psim(t_1),\cdots,\psim(t_m)\}$.
Define $\tpsim_1$ to be the piecewise linear interpolation between the points \allowbreak $\left\{\left(t_i, \psim_{1}\left(t_i\right)\right)\right\}_{i=1}^m$. Let $\tilde{\lambda}_1, \tilde{\lambda}_2,\cdots, \tilde{\lambda}_m$ denote the eigenvalues of $-\frac{1}{2}P\mathcal{D}^{(2)}_{\varphi_m}P$, where $P=I-J/m$, $J\in\RR^{m\times m}$ is the matrix of all ones, and $\left(\mathcal{D}_{\varphi_m}\right)_{i,j}=d_{MV}\left(X_{t_i}^{(m)},X_{t_j}^{(m)}\right)$.

We say the latent position process $X_t$ is \emph{asymptotically Euclidean 1-realizable with asymptotic mirror $\psi_0$} if the following hold:
\begin{enumerate}
\item There exists a function $\psi_0:[0,1]\rightarrow\RR$ and signs $w_m\in\{\pm1\}$  such that $\sup_{t\in[0,1]}|\tpsim_{1}(t)-w_m\psi_0(t)| \to 0 \text{ as } m \to \infty,$
\item As $m\rightarrow\infty$,
$$
\max_{i,j\in\{1,\ldots,m}\bigg|\left(\mathcal{D}_{\varphi_m}\right)_{i,j} -  |\psi_0(t_i)-\psi_0(t_j)|\bigg|\rightarrow0
$$
\item $\tilde{\lambda}_1  \sim \Theta(m), $
\item $\frac{ \sum^m_{i=2} \tilde{\lambda}^2_i  }{ \tilde{\lambda}_1 } \sim O\left(\frac{1}{m}\right).$
\end{enumerate}

\end{definition}
To consider the impact of this denser sampling in time when $m$ grows large, we consider a time rescaling of Model \ref{changepointmodel}. This leads us to a notion of an {\em in-fill random walk latent position process with changepoint}. 
\begin{model}[In-fill random walk LPP with changepoint]
\label{infillmodel}
Let $t^*\in(0,1)$. For any $m\geq 2$, set $c=0$, $\delta_m=1/m$, and let $\tilde{X}_t^{(m)}$ follow the random walk LPP of Model \ref{changepointmodel} with first-order jump probability changepoint at $t_m^*=\lfloor t^* m\rfloor$. Put $t_i = i/m$ for $i\in\{1,\ldots,m\}$ and define the \emph{in-fill random walk LPP with changepoint at $t^*$} as
$$ X_{t_i}^{(m)} = \tilde{X}_i^{(m)}, 1\leq i\leq m. $$
That is, 
\begin{align*}
X^{(m)}_0 &=0 \quad \text{with probability 1}, \notag \\
X^{(m)}_{t_i}&=\begin{cases}
X^{(m)}_{t_{i-1}}+\frac{1}{m} \quad \text{with probability $p_{t_i}$} \\
X^{(m)}_{t_{i-1}} \quad \text{with probability $1-p_{t_i}$},
\end{cases}
\end{align*}
where $p_{t_i} = p$ for $t_i \leq t^*$ and $p_{t_i} = q$ for $t_i > t^*$.
\end{model}

We now present our result describing the behavior of the mirror for the LPP in Model~\ref{infillmodel} as the number of time-points tends to infinity. In particular, a univariate, piecewise linear function with a change in slope at the changepoint approximates the pairwise distances closely, and can be uniformly approximated by the first dimension of CMDS.  Since this asymptotic mirror reveals the exact time of the changepoint, we can use it for consistent changepoint detection once the number of time-points becomes large (see Section~\ref{Sec:changepoint analysis}).

\begin{theorem}
\label{thm:ourmodelisasysum}
Consider the in-fill random walk latent position process in Model \ref{infillmodel} with changepoint at $t^*\in(0,1)$. Define the function $\psi_Z$ as follows:
$$
\psi_Z(t)=
\begin{cases}
    pt+c_0 \mbox{~for~} t \leq t^* \\
    qt+(p-q)t^*+c_0 \mbox{~for~} t > t^*. \\
\end{cases} 
$$
$$
\text{where $c_0=t^*\left(p-q\right)\left(\frac{t^*}{2}-1\right)-\frac{q}{2}$ such that $\int_0^1 \psi_Z(t) dt=0$.}
$$ 
Then this LPP is in-fill asymptotically Euclidean 1-realizable, with asymptotic mirror $\psi_Z$.
\end{theorem}

Now we turn to the problem of changepoint estimation for Model~\ref{infillmodel} and more generally, asymptotically Euclidean 1-realizable LPPs that have a piecewise linear mirror, as we have observed in multiple real-data settings.

\section{Changepoint analysis}
\label{Sec:changepoint analysis}

In Model \ref{infillmodel}, at the changepoint $t^*$, the jump probability changes in the LPP itself. As we defined in Definition \ref{def:highchangepoints}, this is a first-order changepoint for the LPP. Indeed, for a fixed $m$ and any $i \in \{1,2,\cdots,m\}$:
$$ \Delta_{1}^{\frac{1}{m}}(t_i) \overset{\mathcal{L}}{=} \begin{cases} \frac{1}{m}K_p &\text{if }t_i\leq t^*,\\
\frac{1}{m}K_q&\text{if }t^*+\frac{1}{m}<  t_i,\\ \frac{1}{m}K_q&\text{if }t^*< t_i \leq t^*+\frac{1}{m},\end{cases}$$ where $K_p$ is a Bernoulli random variable with success probability $p$.  
In this section, we adapt techniques from \cite{athreya2025euclidean} to show that network realizations from the in-fill LPP model (Model \ref{infillmodel}) can be used to estimate Euclidean mirrors and localize changepoints associated to the underlying latent position process. 

\subsection{Estimation of model parameters from network realizations}\label{sec:mirror estimation}

Our network time series consists of time-indexed graphs on a common vertex set, and for any given time, each vertex has a corresponding time-dependent latent position. The latent positions are typically unknown, but can be consistently estimated through spectral decompositions of the observed adjacency matrices \cite{STFP-2011}
.  Suppose that $G$ is a random dot product graph with latent position matrix $\bX \in \R^{n \times d}$, where the rows of $\bX$ are independent, identically distributed draws from a latent position distribution $F$ on $\mathbb{R}^d$. Let $\bP=\bX \bX^T$ be the connection probability matrix and $\bA$ be the adjacency matrix for this graph. The {\em adjacency spectral embedding} (ASE) is a rank $d$ eigendecomposition of the adjacency matrix.

\begin{definition}[Adjacency Spectral Embedding]
Given an adjacency matrix $\bA$, we define the \emph{adjacency spectral embedding} (ASE) with dimension $d$ as $\hat{\bX}=\hat{\bU}|\hat{\bS}|^{1/2}$, where $\hat{\bS}\in\R^{d\times d}$ is the diagonal matrix with the $d$ largest-magnitude eigenvalues of $\bA$ on its diagonal, arranged in decreasing order. $\hat{\bU}\in\R^{n\times d}$ is the matrix of $d$ corresponding orthonormal eigenvectors arranged in the same order. $\hat{\bX}$ is the estimated latent position matrix.  
\end{definition}

When the true dimension $d$ of the latent positions---or equivalently, the rank of $\bP$---is unknown, we can infer it using the scree plot \cite{zhu2006automatic}. The ASEs of the observed adjacency matrices in our TSG  at times $t$ and $s$, denoted $\hat{\bX}_t$ and $\hat{\bX}_s$, are estimates of the latent position matrices $\bX_t$ and $\bX_s$, from which we obtain an estimate $\hat{d}_{MV}$, defined below, of the $d_{MV}$ distance between the latent position random variables over time. 
\begin{definition}\label{def:hat_dmv}
The {\em estimated} $d_{MV}$ distance $\hat{d}_{MV}$ is defined as 
\begin{equation}\label{equ:hat_dmv}
\hat{d}_{MV}(\hat{\bX}_{t},\hat{\bX}_{s}):=\min_{W\in\mathcal{O}^{d\times d}} \frac{1}{\sqrt{n}}\|\hat{\bX}_{t}-\hat{\bX}_{s}W\|_2.
\end{equation}
\end{definition}
In\cite{athreya2025euclidean}, the authors show that $\hat{d}_{MV}$ is a consistent estimate for the true dissimilarity. 
\begin{theorem}
\label{thm:approximation}\cite{athreya2025euclidean}
With overwhelming probability,
$$
\left|\hat{d}_{MV}(\hat{\bX}_{t},\hat{\bX}_{s})^2-d_{MV}(\varphi(t),\varphi(s))^2\right|\leq \frac{\log(n)}{\sqrt{n}}.
$$
\end{theorem}
Sampling our network time series at $m$ time points, we obtain $m$ adjacency matrices whose spectral embeddings yield the matrix $\widehat{\mathcal{D}}$ of pairwise estimated dissimilarities:
$$
\left(\widehat{\mathcal{D}}_{\varphi}\right)_{s,t}=\hat{d}_{MV}(\hat{\bX}_{s},\hat{\bX}_t).
$$
As in Definition \ref{def:CMDS}, let $P:=I_m-\frac{J_m}{m}$ where $I_m$ is the $m \times m$ identity matrix and $J_m$ is the $m \times m$ all ones matrix. Recall from Definition \ref{def:0-skeleton-dMV-mirror} that the $c$-dimensional zero-skeleton mirror $\psi$ for the latent position process $\varphi$ is given by the rows of $US^{1/2}$, where $S$ is the diagonal matrix of the $c$ largest positive eigenvalues of $-\frac{1}{2}P\mathcal{D}_{\varphi}^{(2)} P^\top$ and $U$ the matrix of associated eigenvectors. Since $\widehat{D}_{\varphi}$ serves as a reasonable estimate for $\mathcal{D}$, we apply classical multidimensional scaling of this estimated dissimilarity matrix to generate an estimated zero-skeleton mirror.

\begin{definition}[Estimated zero-skeleton mirror]\label{def:est_zero_skeleton_mirror}
Let $\hat{\lambda}_1, \cdots, \hat{\lambda}_c$ denote the $c$ largest eigenvalues, in decreasing order, of $\widehat{\mathcal{D}}_{\varphi}$. Put $\hat{S}_{ii}=\max\{\lambda_i, 0\}$. Let $\hat{U}$ be the ordered matrix of eigenvectors associated to the $c$ largest eigenvalues of $\widehat{\mathcal{D}}_{\varphi}$. Let $\hat{\psi}(t_k)$ denote the $k$th row of $\hat{U}\hat{S}^{1/2}$. The {\em estimated $c$-dimensional zero skeleton mirror} is given by $\{\hat{\psi}(t_1), \cdots, \hat{\psi}(t_m)\}$. 
\end{definition}

The following theorem from \cite{athreya2025euclidean} allows us to control the error between the true and estimated zero-skeleton mirror.

\begin{theorem}[Control between mirror and estimated mirror]
\label{thm:consistency_of_mirror}
Suppose $\mathcal{D}_{\varphi}$ is approximately $c$ Euclidean realizable. Let $\hat{U},U\in\RR^{m\times c}$ be the top $c$ eigenvectors, and 

$\hat{S},S\in\RR^{c\times c}$ be the diagonal matrices with diagonal entries equal to the top $c$ eigenvalues of $-\frac{1}{2}P\widehat{\mathcal{D}}_{\varphi}^{(2)} P^\top$ and $E_{\varphi}:=-\frac{1}{2}P \mathcal{D}_{\varphi}^{(2)} P^\top$, respectively. Suppose $S_{i,i}>0$ for $1\leq i\leq c$. Then with overwhelming probability, there is a real orthogonal matrix $R\in\mathcal{O}^{c\times c}$ such that 
$$\|\hat{U}-UR\|_F\leq \frac{2^{3/2}}{\lambda_c(E_{\varphi})}\left(\frac{m\log(n)}{\sqrt{n}}+\left(\sum_{i=c+1}^{m}\lambda^2_i\left(E_{\varphi}\right)\right)^{1/2}\right).$$ Call this upper bound $B=B(n,m,c)$. The CMDS output satisfies $$\|\hat{U}\hat{S}^{1/2}-US^{1/2}R\|_F\leq B\lambda_1^{1/2}(E_{\varphi})\left(2+4B\kappa^{1/2}+(1+2B)\frac{m\log(n)}{\sqrt{n}\lambda_c(E_{\varphi})}\right),$$
where $\kappa=\lambda_1(E_{\varphi})/\lambda_c(E_{\varphi})$. In particular, we have
$$\sum_{i=1}^{m} \|\hat{\psi}(t_i)-R\psi(t_i)\|^2\leq B^2\lambda_1(E_{\varphi})\left(2+4B\kappa^{1/2}+(1+2B)\frac{m\log(n)}{\sqrt{n}\lambda_c(E_{\varphi})}\right)^2.$$

\end{theorem}

We can simplify the upper bound by further assuming that $c=1$ and $\lambda_1(E_{\varphi}) \sim \Theta(m)$, this gives the following corollary. 

\begin{corollary}\label{cor:simple_upperbound}
If additionally we require $c=1$ and $\lambda_1(E_{\varphi}) \sim \Theta(m)$, then with overwhelming probability, there is a $r \in \{\pm1\}$ and a constant $C$
such that
$$ \sum_{i=1}^m |\hat{\psi}(t_i)-r\psi(t_i)|^2\leq \frac{C}{\lambda_1(E_{\varphi})}\left(\frac{m\log(n)}{\sqrt{n}}+\sqrt{\sum_{i=2}^m \lambda_i^2(E_{\varphi})} \right)^2.$$ In particular,
\begin{align*}
\max_{1\leq i \leq m} |\hat{\psi}(t_i)-r\psi(t_i)| \leq  C \left( \frac{m \log(n)}{\sqrt{n} \sqrt{\lambda_1(E_{\varphi})}} +  \sqrt{\frac{\sum^m_{i=2} 
\lambda^2_i\left(E_{\varphi}\right)}{\lambda_1 \left(E_{\varphi}\right)}}  \right).
\end{align*}
\end{corollary}

\begin{figure}
    \centering
    \includegraphics[width=4.04in]{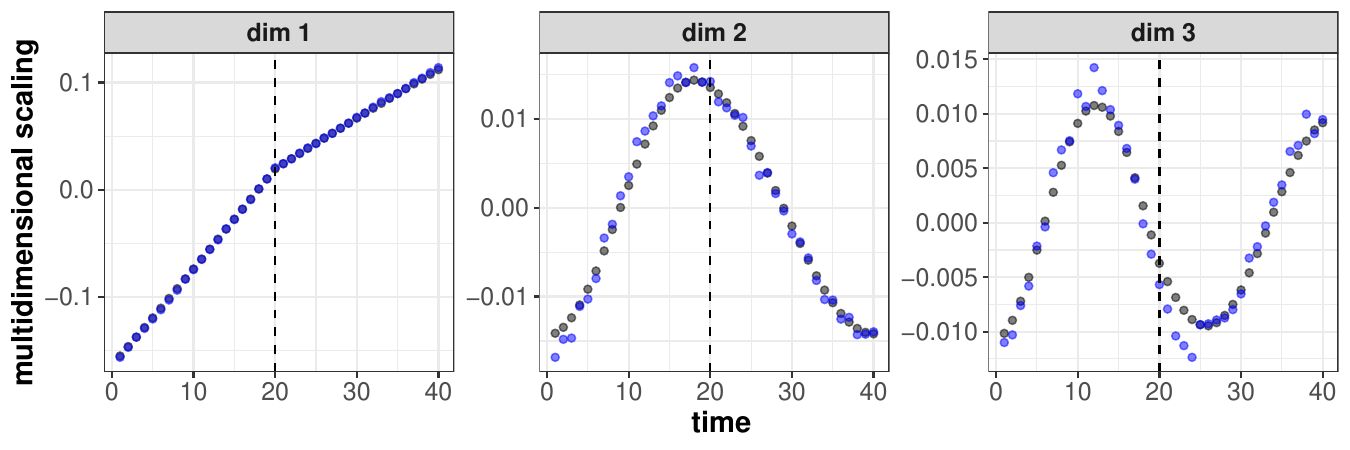}
    \caption{CMDS results on first 3 dimensions for Model \ref{changepointmodel} with same setting as in Figure \ref{fig:cmds_changepoint}.
    The black dots are the numerical CMDS result on $\mathcal{D}^{(2)}_{\varphi_m}$. The blue dots are estimation from one realization of time series of graphs with $n=1500$. Blue dots precisely aligns with the black dots in the first dimension. In the second dimension, the alignment of the estimated mirrors is less accurate, and the discrepancy increases further in the third dimension. For more details, see the final section of the Appendix.
    }  
    \label{fig:cmds_changepoint_with_blut_dots}
\end{figure}

The results above guarantee that for a fixed number of graphs $m$ in the TSG, as the number of vertices tends to infinity, at time $t_i$, the estimated mirror $\hat{\psi}(t_i)$ will be close to the true mirror $\psi(t_i)$ in the low dimensional Euclidean space. This is depicted in Figure \ref{fig:cmds_changepoint_with_blut_dots}, where, with $n=1500$, the estimated mirror (blue dots) precisely aligns with the true mirror (black dots) in the first dimension. In the second dimension, the alignment of the estimated mirrors is less accurate, and the discrepancy increases further in the third dimension.

We exploit these results to show the consistency of localizing the changepoint using TSG realizations in the Section \ref{Sec:consistency}.

\subsection{Consistency of the localizing estimator}\label{Sec:consistency}

We have seen that the latent position process of Model~\ref{infillmodel} is asymptotically Euclidean 1-realizable, and that its first-order changepoint $t^*$ is reflected in the asymptotic mirror by a change in slope at this point. Thus, an estimator of the location of this slope change may be useful for localizing the underlying changepoint in the latent position process.  We establish this in the following theorem, where we show that for any asymptotically 1-realizable latent position process whose asymptotic mirror is piecewise linear with a slope change, the location of this slope change can be consistently estimated. Our estimator for the changepoint is given by the point of slope change for the closest piecewise-linear continuous function with a single change in slope at one of the observed times to the estimated mirror (see Algorithm~\ref{alg:1}). 

\begin{theorem}\label{thm:consistency asy1-d}
Consider an asymptotically Euclidean 1-realizable latent position process with zero-skeleton mirror $\left\{\psim\left(t_i\right),1\leq i \leq m\right\}$ and asymptotic mirror $\psi_0(t) = at+b+\Delta(t-t^*)I_{\{t>t^*\} }$, where $ t \in [0,T]$ and $\Delta\neq0$. Denote by $\tpsim_1$ the piecewise linear interpolation between the points
$$\left\{\left(t_i, \psim_{1}\left(t_i\right)\right) | 1 \leq i \leq m \right\} \subset \mathbb{R} \times \mathbb{R}.$$ 
Consider the time series of networks on $n$ vertices at points $\mathcal{T}=\{t_1, \cdots, t_{m}\}$ with this latent position process. 
Put $E_{\psim}=-\frac{1}{2}P\mathcal{D}_{\psim}^{(2)}P$, and $\mathcal{D}_{\psim}^{(2)}$ is the $m \times m$ entry-wise square of the distance matrix of the mirror $\psim$. Assume $t_1=0$, $t_{m}=T$, $t^* \in (t_1,t_{m})$ and $\lambda_1 \left(E_{\psim}\right) \sim \Theta(m)$.
Let $\left\{\hpsim_1(t_i) | 1\leq i \leq m \right \}$ denote the spectrally-derived estimated first dimension mirror of 
\cite{athreya2025euclidean}. 
Let $\rho=\mathop{max}\limits_{1\leq i < m} |t_i-t_{i+1}|$.
Let $\hat{t}$ be the $l_{\infty}$ localization estimator defined by:
\begin{equation}\label{eq:l-infchangepoint_def}
\adjustbox{width=4.513in}{$\displaystyle \hat{t},\hat{\alpha},\hat{\beta}_L,\hat{\beta}_R=\underset{\substack{t'\in\mathcal{T}\\ \alpha, \beta_L, \beta_R, \beta_L\neq \beta_R}}{\mathrm{argmin}}
\max_{t \in \mathcal{T} } \left| \alpha + \beta_L(t-t')+(\beta_R-\beta_L)(t-t')I_{\{t > t'\}} - \hat{\psi}(t)  \right|. $}
\end{equation}
Then for $m$ and $n$ sufficiently large, there exists a constant $C$ and a $w_m\in\{\pm1\}$ such that with high probability:\\
\adjustbox{width=4.75in}{
$\displaystyle
|\hat{t}-t^*| \leq \left( 
\frac{C m \log(n)}{\sqrt{n} \sqrt{\lambda_1 \left(E_{\psim}\right)} |\Delta|} + \frac{ C \sqrt{\sum^m_{i=2} \lambda^2_i\left(E_{\psim}\right)} }{ \sqrt{\lambda_1 \left(E_{\psim}\right)} |\Delta|} + \frac{4 \sup_{ t \in [0,T] } \left|\tpsim_1(t)- w_m\psi_0(t)\right| }{|\Delta|} + 2\rho
\right) \frac{T}{ min\{T-t^*,t^*\} } .
$}
\end{theorem}

\begin{remark}
We can interpret the terms in this bound as follows. The first term captures the statistical error in estimating the matrix of pairwise distances from observed networks; the second term captures the distortion in estimating the matrix of true pairwise distances using a one-dimensional Euclidean distance matrix. The third term captures the distortion in the true first dimension of the mirror arising from the in-fill asymptotics: as $m$ grows, we know that the first dimension of the mirror looks like the piecewise-linear function $\psi_0$, but for finite $m$, $\tilde{\psi}_1^{(m)}$ will still not be exactly equal to this function.The last term inside the parentheses captures the mesh of the sampled points, which can lead to an error in estimating $t^*$ since our estimate $\hat{t}$ only takes values on observed values of $t$. Finally, the ratio multiplying the terms in parentheses captures the relative difficulty of estimating changepoints near the ends of the observed interval compared to the center of the interval. When the changepoint is near the center of the interval, there is much more information about the drift in the process before and after the changepoint than in the case that the changepoint happens near one of the two endpoints.
\end{remark}

This bound can be simplified significantly when we have exact (non-asymptotic) Euclidean 1-realizability, and the sampled times are uniformly spaced.

\begin{corollary}\label{Cor:uniform time step result}
Consider the setting of Theorem~\ref{thm:consistency asy1-d}, where the LPP is exactly Euclidean 1-realizable, and we uniformly sample the time points in [0,T]: that is, for any $1 \leq i < m$, $t_{i+1}-t_i=\frac{T}{m-1}$. Then for large $m$ and $n$, there exists a constant $C$ such that with high probability: 
$$
|\hat{t}-t^*| \leq  \left( \frac{C \log(n) \sqrt{mT} }{\sqrt{a^2 T^3+\Delta^2(T-t^*)^{3}} |\Delta| \sqrt{n} }   + \frac{6T}{m-1} \right) \frac{T}{ min\{T-t^*,t^*\}}. 
$$      
\end{corollary}

\begin{algorithm}[t]
\caption{$l_{\infty}$ localization for piecewise linear data}
\label{alg:2}
\begin{algorithmic}[1]
\State Input: $m$ pairs of observations $\{(t_1,y_1),(t_2,y_2),\cdots,(t_m,y_m)\}$ with each $t_i\in \R$,$y_i\in \R$.
\State \textbf{for k from $2$ to $m-1$:}
\State \quad Define $$S_k:=\min_{\alpha, \beta_L, \beta_R \in \R, \beta_L \neq \beta_R } 
\max_{ 1\leq i \leq m } \left| \alpha + \beta_L(t_i-t_k)+(\beta_R-\beta_L)(t_i-t_k)I_{\{t_i > t_k\}} - y_i \right|.$$
\State Find the smallest $k_0$ such that $S_{k_0}= \min\{S_2,S_3,\cdots,S_{m-1}\}$ and set $\hat{t}=t_{k_0}$.
\State Output: $\hat{t}$. 
\end{algorithmic}
\end{algorithm}

For consistent estimation of $t^*$, we apply the high-probability finite-sample bound from Theorem~\ref{thm:consistency asy1-d} with some additional assumptions on the growth of $m$ versus $n$, which gives the following corollary:

\begin{corollary}[Consistency of $\hat{t}$ for in-fill asymptotically 1-d Euclidean realizable with piecewise linear asymptotic mirror]\label{Cor:asy-sum-consistency}
Consider the setting of Theorem~\ref{thm:consistency asy1-d}. 
Suppose that when $m \to \infty$, $\rho_m=\max_{1 \leq i \leq m-1} |t_i-t_{i+1}|  \to 0$. Suppose also that when $n\to \infty$ and $m \to \infty$, $\frac{\log(n)\sqrt{m}}{\sqrt{n}} \to 0$.
Then for the $l_\infty$ estimator $\hat{t}$ defined as above, we have with high probability that 
$$
|\hat{t}-t^*| \to 0 \text{ as } m \to \infty, n \to \infty.
$$  
\end{corollary}

In particular, these results apply to LPPTSGs coming from Model~\ref{infillmodel}, so we have consistent localization for this model as long as $m$ does not grow too fast relative to $n$.

\begin{corollary}\label{Cor:3}
Suppose $n, m$ are such that $\frac{\log(n)\sqrt{m}}{\sqrt{n}} \to 0$ as $n, m \rightarrow \infty$. Using the $l_\infty$ localization estimator on the observed time series of networks generated from Model \ref{infillmodel}, we can consistently localize $t^*$ as $n \to \infty$ and $m \to \infty$. 
\end{corollary}

\subsection{Changepoint detection}
\label{sec:detection}

So far, we have considered the problem of changepoint \emph{localization}, where we assume the existence of the changepoint to be known, and seek to estimate its location. However, another important problem is detection itself. Our changepoint localization estimator gives rise to a test statistic for the changepoint detection problem: we can look at the best-fitting piecewise-linear approximation to the first dimension of the Euclidean mirror, and use the change in slope, $|\hat{\beta}_R-\hat{\beta}_L|$, between the left and right linear pieces of that piecewise linear approximation. In general, formulating a null model under which the distribution of this test statistic can be studied is a very challenging problem, but we can consider the special case of Model~\ref{infillmodel} to gain insight into its performance. We simulating LPPTSGs coming from this model under the null hypothesis, where there is no change in slope, in order to obtain a bootstrap estimate of the distribution of this test statistic. Then for a realized LPPTSG coming from Model~\ref{infillmodel} with a changepoint, we obtain an estimated $p$-value, and we can approximate the power of this test if we specify a level of Type 1 error. 

\begin{figure}
    \begin{subfigure}{0.49\textwidth}
        \includegraphics[width=\textwidth]{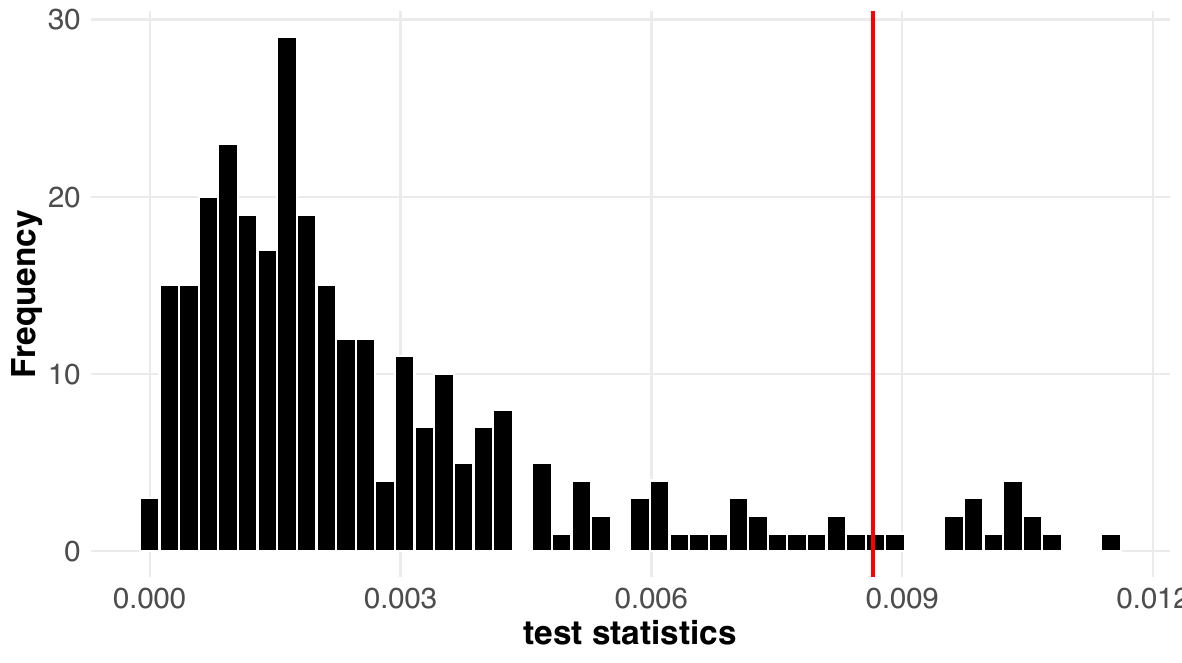}
        \caption{Distribution of $|\hat{\beta}_R-\hat{\beta}_L|$ under the null hypothesis $p=q=0.4$}
    \end{subfigure}
    \hfill
    \begin{subfigure}{0.49\textwidth}
        \includegraphics[width=\textwidth]{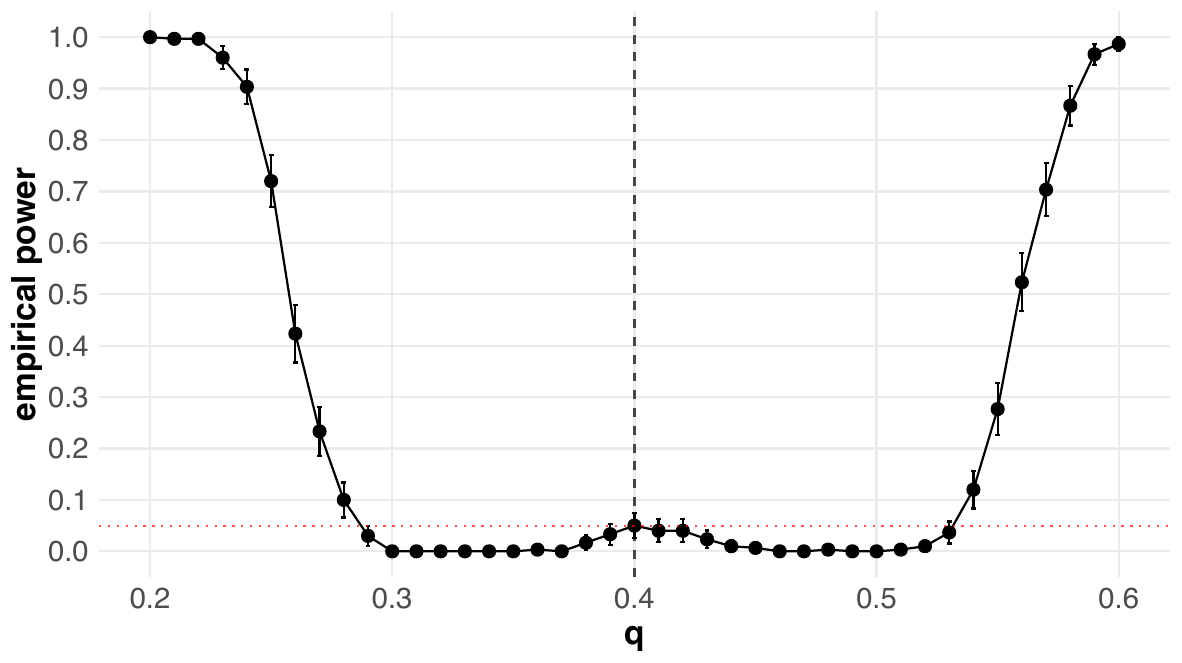}
        \caption{Estimated power of $|\hat{\beta}_R-\hat{\beta}_L|$ against various alternatives}
    \end{subfigure}
    \caption{Bootstrapped critical value calculation based on $nmc = 300$ and estimated power for the changepoint detection problem, in the case that $n=500$, $m=16$, $p=0.4$ also with $nmc = 300$ for each pair of $(p,q)$. We see that this test statistic achieves high power once the change in slope is significant enough.}
    \label{fig:changepointdetection}
\end{figure}

In Figure~\ref{fig:changepointdetection} we show the results of this approach, in a setting with $n=500$, $m=16$, and $p=0.4$. Under the null hypothesis, we see that the density of our test statistic decays quickly, exhibiting similar behavior to a generalized extreme value distribution, in keeping with this estimator measuring a maximum among errors in the mirror estimation. Under the null hypothesis, there is no change in slope in the true mirror, so the distribution of our test statistic incorporates both randomness in the choice of $\hat{t}$ as well as in estimating the slopes for both linear pieces. Under the alternative, there is a change in slope for the true mirror, so we expect the randomness in our change of slope estimator to mostly arise from the estimation of the slopes for the left and right linear pieces. As such, the mean of the distribution of the test statistic becomes nonzero for weak alternatives, but the variance decreases significantly compared to the null hypothesis, leading to a loss of power. Once the alternative becomes strong enough, the mean of the distribution of the test statistic becomes more nonzero, and coupled with the small variance, we see a rapid increase in power.

\subsection{Numerical experiments}
\label{sec:numerical}

We begin by investigating the performance of our changepoint localization estimator for LPPTSGs arising from Model~\ref{infillmodel} as $n$ and $m$ vary.\footnote{All of our codes are available: \url{https://github.com/TianyiChen97/Euclidean-mirrors-and-first-order-changepoints-in-network-time-series}} %

For fixed values of $n$, $m$, $p=0.4$, $q=0.3$, and changepoint $t_m^*=\frac{m}{2}$, we generate a time series of graphs as follows. We fix a vector $X_{0} \in \R^{n}$ with all entries set to $0.1$, then for each time $t$, $X_t=X_0+\sum_{s=1}^t \frac{0.9}{m}U_s$, where $U_s\in\RR^n$ has entries $U_{is}\sim\mathrm{Bernoulli}(p_t)$, which are independent for all $i$ and $s$. The jump probabilities $p_t=p$ for $t\leq t_m^*$ and $p_t=q$ for $t> t_m^*$. Then for each time, we generate an RDPG using the latent positions in $X_t$, obtaining a TSG $\{A_{1},\cdots,A_{m}\}$ as defined in Definition \ref{def:tsg}.

For mirror estimation, we compute the covariance-based dissimilarity Eq (\ref{equ:hat_dmv}) between pairs of adjacency matrices $A_{k}, A_{\ell}$ in  $\{A_{1},\cdots,A_{m}\}$; then perform CMDS with embedding dimension $d=1$ (note that in this case, further computation of ISOMAP is unnecessary).
We then compute the $l_\infty$ localization estimator for the changepoint, following Algorithm \ref{alg:2}. In Algorithm \ref{alg:2}, the optimization at Step 3 is reformulated as a constrained linear programming problem, which we solve using the lpSolveAPI \footnote{\url{https://cran.r-project.org/web/packages/lpSolveAPI/index.html}} function in R. This procedure is repeated for each Monte Carlo simulation, with each iteration yielding $\hat{t}_{mc}$.

\begin{figure}[t]
     \centering
     \subfloat{\includegraphics[width=2.375in]{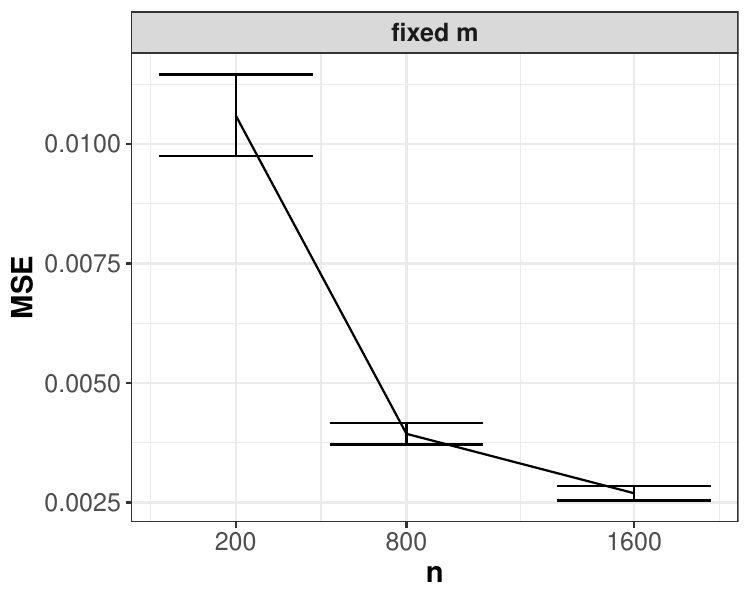}}
     \subfloat{\includegraphics[width=2.375in]{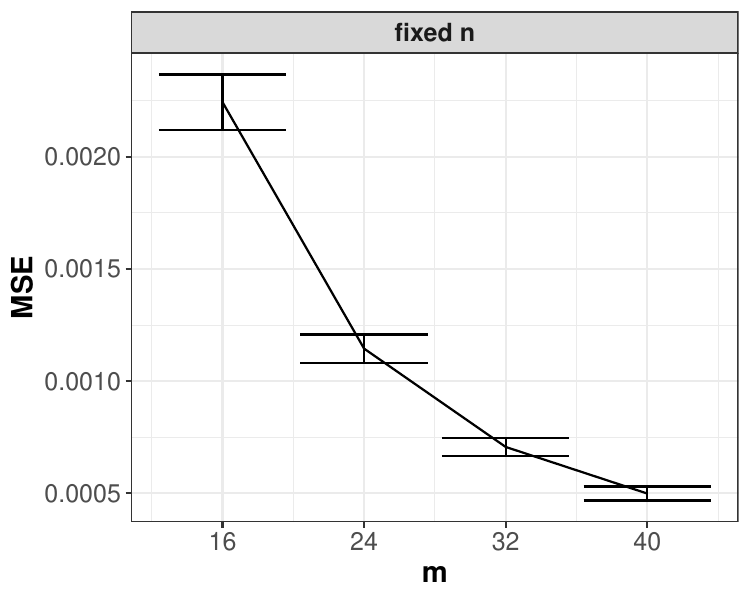}}
    \caption{
   Accuracy of network changepoint estimator $\widehat{t}$ as a function of $m$ and $n$ when using the 1st dimension of the estimated mirror. $p=0.4$, $q=0.3$, $|p-q|=0.1$, $nmc=2000$ and $t^* = \frac{1}{2}$ fixed for both figures. Left: $m=12$ fixed. Right: $n=800$ fixed. We see the MSE tends to 0 for fixed $n=800$ with $m$ growing and for fixed $m=12$ with $n$ growing.}
\label{fig-changemn}
\end{figure}

To evaluate the performance of our estimator, we consider the squared error $(\hat{t}_{mc} - t^*)^2$ for each iteration. We run the Monte Carlo simulation $nmc = 2000$ times, and record the mean of these 2000 realizations as the mean square error (MSE):
$$
\text{MSE} := \frac{1}{nmc} \sum_{mc=1}^{nmc} \left(\hat{t}_{mc} - t^*\right)^2.
$$Next, we calculate the sample standard deviation for the 2000 realizations: $$
\text{std} := \sqrt{\frac{1}{nmc-1} \sum_{mc=1}^{nmc} \left( \left(\hat{t}_{mc} - t^*\right)^2 - \text{MSE} \right)^2}.
$$ Since we have a large sample from the distribution of $(\hat{t}-t^*)^2$ and we are estimating the mean of this distribution, we may construct a confidence interval for the MSE using the normal approximation as:$$
\text{CI} = \text{MSE} \pm \frac{1.96}{\sqrt{nmc}} \times \text{std}.
$$ 
We consider fixed $n=800$ and vary $m$ over $16,24,32,40$; next, we fix $m=12$ and vary $n$ over $200,800,1600$. The results of these experiments can be seen in Figure \ref{fig-changemn}. For fixed $m$, MSE drops quickly with increasing $n$. For fixed $n=800$, we also see that larger $m$ is associated with significantly smaller MSE.

We have shown that for large $n$ and $m$, the first dimension of the estimated mirror should yield consistent estimates for the changepoint $t^*$. In practice, however, there may still be useful signal contained in CMDS dimensions beyond the first (see Figure \ref{fig:cmds_changepoint}). We now consider computing the estimated mirror $\hat{\psi}$ in $d$ dimensions, before considering the ISOMAP embedding of this curve into one dimension, denoted by $\hat{\psi}_{d\rightarrow1}$. This embedding retains the piecewise linear structure, but incorporates additional signal from other dimensions. We write this iso-mirror via
\begin{equation}
\label{eq:error model}
\hat{\psi}_{d\to1}(t)=\psi_Z(t)+\epsilon_t  \quad t \in [0,1],
\end{equation}
where $\psi_Z$ is defined for some $t^*\in(0,1), p,q>0, p\neq q$ as
$$
\psi_Z(t)=
\begin{cases}
pt &\text{for }t < t^* \\
qt+(p-q)t^* &\text{for }t\geq t^*.
\end{cases}
$$

\begin{remark}
We emphasize that two sources of error lie within the $\{\epsilon_t\}$: the first from finite network size $n$ and the second from the finite number of sampled networks or time-points $m$. The first source of error, addressed in \cite{athreya2025euclidean}, arises because the draws from the LPP are random, as are the observed adjacency matrices $A_t$, making the estimated distances $\min_W\|\hat{X}_t-\hat{X}_{s}W\|/\sqrt{n}$ a noisy approximation of the true dissimilarities $d_{MV}(\varphi_m(t),\varphi_m(s))$. 
The second source of error is from the finite time-sampling of $m$ networks. For finite $m$, the first dimension of the mirror of Model \ref{changepointmodel} is not equal to $\psi_Z$, but instead converges to $\psi_Z$ as $m \to \infty$. This error is deterministic and small for large $m$.
Moreover, since we apply ISOMAP to an estimated mirror, these two sources of error in the mirror estimation propogate into the iso-mirror estimate $\hat{\psi}_{d\to 1}$ as well.
\end{remark}

\begin{algorithm}
\caption{Iso-mirror estimation}
\label{alg:1}
\begin{algorithmic}[1]
\State Input: TSG $\{A_1,A_2,\cdots,A_m\}$, ASE dimension $d_1$, CMDS dimension $d$.
\State Compute $\hat{X}_t=$ASE$(A_t)\in \RR^{n\times d_1}, 1\leq t\leq m$.
\State Construct the distance matrix $\hat{\mathcal{D}}\in\R^{m\times m}$ where $\hat{\mathcal{D}}_{i,j}=\min_{W\in\mathcal{O}^{d_1\times d_1}} \frac{1}{\sqrt{n}}\|\hat{\bX}_{t}-\hat{\bX}_{s}W\|_2$.
\State Compute CMDS$(\hat{\mathcal{D}})=\{\hat{\psi}(1),\hat{\psi}(2),\cdots,\hat{\psi}(m)\}$, with each $\hat{\psi}(t) \in \R^d$. 
\State Apply ISOMAP on $\{\hat{\psi}(1),\hat{\psi}(2),\cdots,\hat{\psi}(m)\}$ using $k$ nearest neighbors, where $k$ is the smallest value such that the $k$-nearest-neighbor graph is connected, obtaining $\{\hat{\psi}_{d\to1}(1),\hat{\psi}_{d\to1}(2),\cdots,\hat{\psi}_{d\to1}(m)\}$, with each $\hat{\psi}_{d\to 1}(t)\in \RR$.
\State Output: $\{\hat{\psi}_{d\to1}(1),\hat{\psi}_{d\to1}(2),\cdots,\hat{\psi}_{d\to1}(m)\}$.
\end{algorithmic}
\end{algorithm}

\begin{figure}
     \centering
     \includegraphics[scale=0.45]{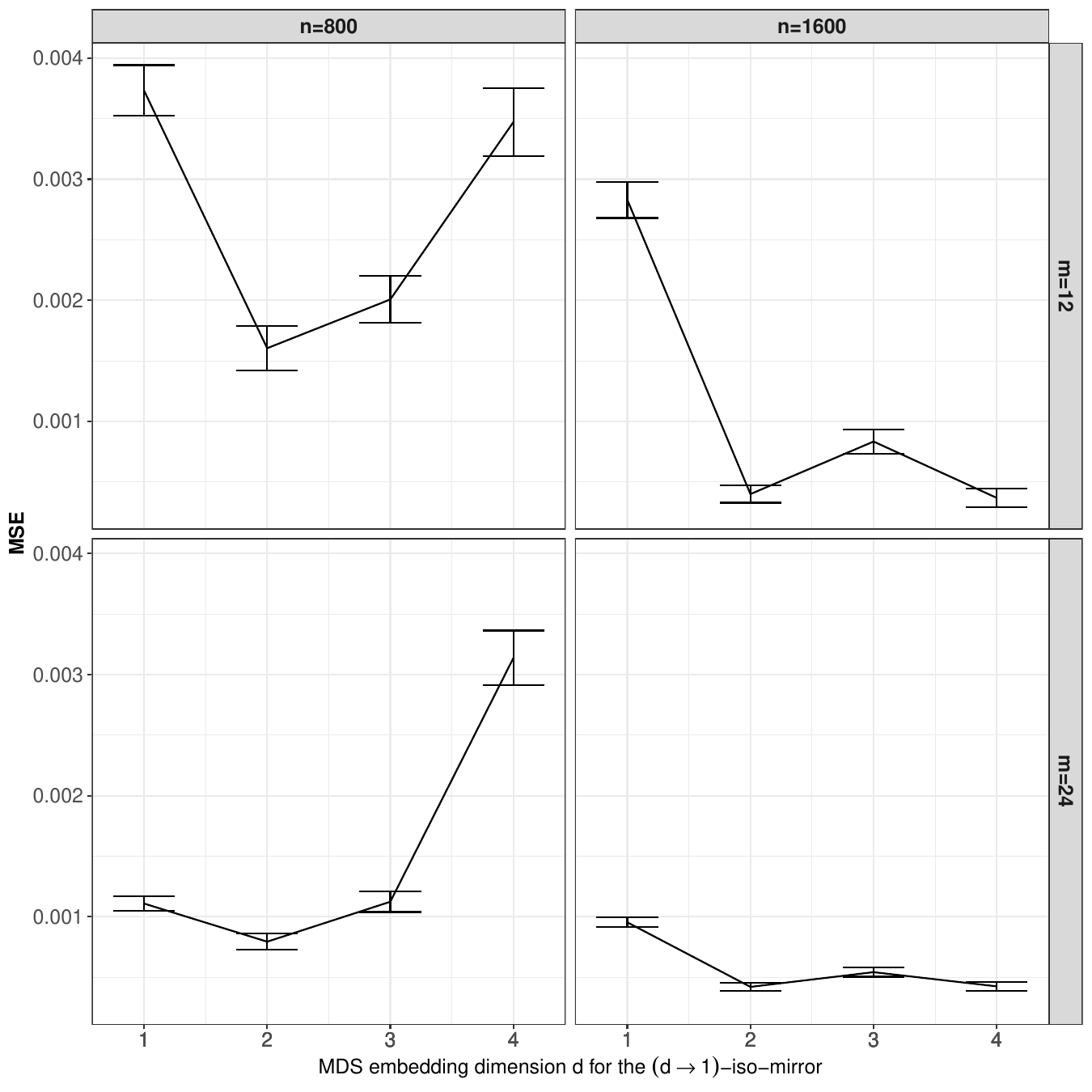}
    \caption{
Accuracy of network changepoint estimator $\widehat{t}$ as a function of CMDS embedding dimension $d$ when using the $(d \to 1)$-iso-mirror $\widehat{\psi}_{d \to 1}$. Upper left: $n=800$, $m=12$. Upper right: $n=1600$,$m=12$. Lower left: $n=800$, $m=24$. Lower right: $n=1600$, $m=24$.
In all cases, $p=0.4$, $q=0.3$, $t^* = \frac{1}{2}$ and $nmc=2000$.
Comparing left two panels and right two panels, we see as $n$ increases, the $l_{\infty}$ estimator can yield improvement at higher embedding dimensions. Comparing the upper two panels and the lower two panels, we see the diminishing returns as to higher-dimensional embedding; as $m$ increases, these higher dimensions contain less signal. 
        }
    \label{fig-trunk}
\end{figure}

We examine how the accuracy of our localization estimator $\hat{t}$ depends on the CMDS embedding dimension $d$ for the iso-mirror. All of the details of the network realizations are the same as the first set of experiments, but we now apply our changepoint localization estimator to the iso-mirror estimate $\hat{\psi}_{d\to1}$ rather than the first dimension of the mirror obtained via CMDS. An algorithm for finding the iso-mirror is given in Algorithm~\ref{alg:2}. The results are shown in Figure \ref{fig-trunk}. We see that as the embedding dimension $d$ increases, the MSE first decreases and then, eventually, increases. In our model, higher dimensions still contain changepoint-relevant signal, so some number of additional dimensions can improve localization by reducing bias.  However, embedding into significantly higher dimensions can also introduce additional variance, exhibited by the characteristic $U$ shape of the MSE. 
Comparing the left two panels and the right two panels, we see as $n$ increases, the variance drops dramatically, so increasing the dimension can improve estimation accuracy. Comparing the upper two panels and the lower two panels, we see the decreasing returns of increasing dimension: because of approximate one-dimensional Euclidean realizability, as $m$ increases, there is less signal in the higher dimensions. 

We also compare our mirror-based estimation approach to a simpler method based on fitting a piecewise-linear curve to marginal network statistics at each time, specifically Leiden modularity, Louvain modularity, and the square root of the average degree, since on average, this quantity exhibits a piecewise-linear structure for the Model \ref{changepointmodel}. We see that our iso-mirror based methods perform well compared to modularity-based approaches for these networks, but they are beaten by average degree, since the Model \ref{changepointmodel} has increasing trajectories for the latent positions, and the rate of that increase provides information about the changepoint-- this signal is captured very effectively by the average degree statistic. However, in the next section we discuss time series of brain organoid networks, where our mirror-based approach leads to better estimation of a changepoint compared to average degree and other summary statistics (shown in Appendix~\ref{sec:organoiddetails}).

\begin{table}[htbp]
    \begin{tabular}{lccc}
        \hline
        \textbf{Method} & \textbf{Lower Bound} & \textbf{Mean} & \textbf{Upper Bound} \\
        \hline
        $(d =1 \to 1)$-iso-mirror   & 0.0022  & 0.0024  & 0.0027  \\
        $(d =2 \to 1)$-iso-mirror    & 0.0010  & 0.0013  & 0.0015  \\
        $(d =3 \to 1)$-iso-mirror    & 0.0013  & 0.0015  & 0.0018  \\
        Leiden Modularity  & 0.035   & 0.035   & 0.036   \\
        Louvain Modularity & 0.032   & 0.033   & 0.034   \\
        $\sqrt{\text{Average Degree}}$     & 0.00023 & 0.00033 & 0.00042 \\
        \hline
    \end{tabular}
    \caption{Comparison of MSEs across representations using $\ell_\infty$ localizer for $p=0.4$, $q=0.3$, $n=800$, $m = 16$, $t^* = 8$. We use $nmc = 500$ Monte Carlo replicates for determining the mean, and obtain the upper and lower bounds for the confidence interval for the MSE based on the normal approximation described above. The square root of the average degree in Model \ref{changepointmodel} can be shown to be piecewise linear, which makes it highly compatible with our estimator that finds a best-fitting piecewise linear curve. Network modularity is reported at each time as the maximum value of the objective function obtained via the Leiden and Louvain algorithms. We note that for this problem, random guessing results in an MSE of 0.064, so all methods are detecting some signal.
    }
    \label{tab:mirror_compare_avg_deg_MSE}
\end{table}

\subsection{Analysis of brain organoid data}

In this section, we apply our mirror-based techniques to real data, specifically a time series of brain organoid connectomes \cite{puppo2021super}. Brain organoids are self-organizing structures composed of roughly 2.5 million neural cells, generated from human induced pluripotent stem cells (hiPSCs) \cite{muguruma2015self}. 
To characterize the functional development of the organoids, extracellular spontaneous electrical activity is recorded approximately weekly.
Each time series consists of five minutes of recorded neural activity across 10 months. To approximate the functional connections between neurons, we apply the algorithm proposed in \cite{puppo2021super} and force the accepted degree of temporal approximation to be 0. 

\begin{figure}
    \centering
    \subfloat{\includegraphics[width=2.375in]{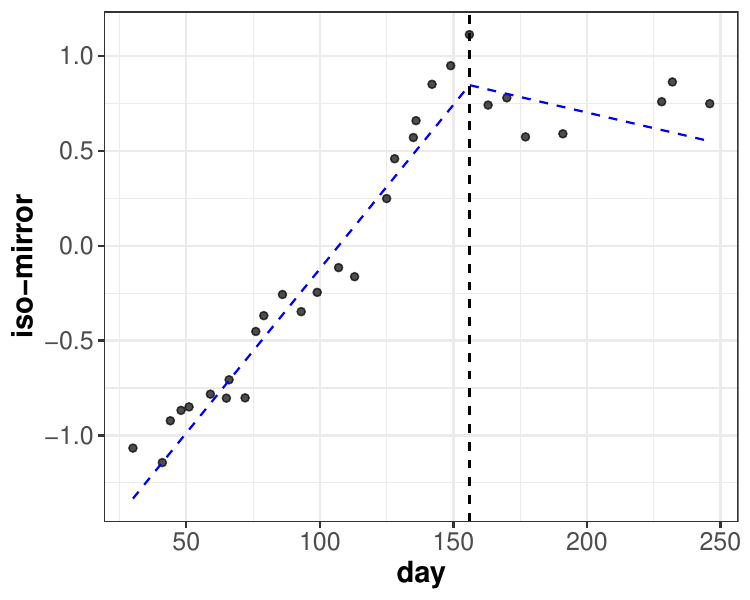} }
    \subfloat{\includegraphics[width=2.375in]{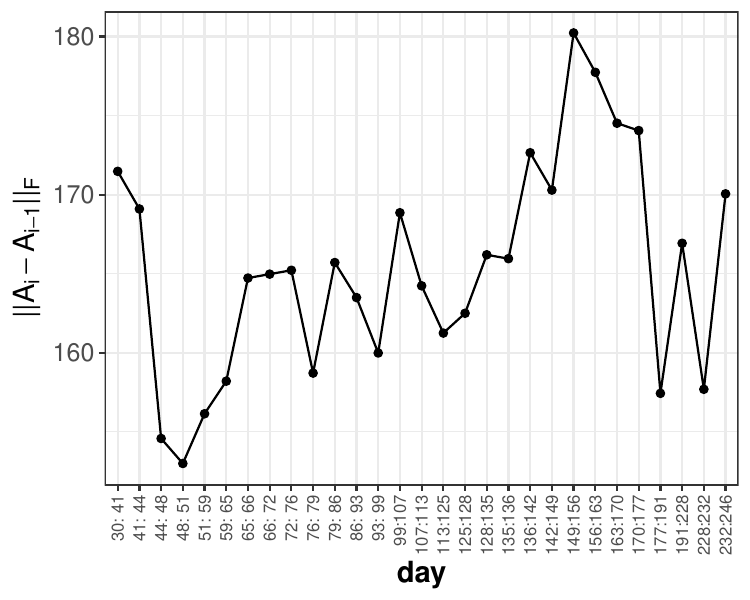}}
    \caption{Real data example with time series of approximate functional connectivity graphs.
    On the left panel is our iso-mirror result. The black dots are the estimated iso-mirror. The dashed blue piecewise linear curve gives the best $l_\infty$ regression fit, with the vertical line as the estimated changepoint on day 156.
    In the right panel, we show the control chart for pairwise Frobenius norm differences for the graphs. It also shows $\|A_{156}-A_{149}\|_F$ is the maximum. However the chart indicates the time series of networks are constantly changing and the underlying dynamic of the networks cannot be described as a zeroth-order changepoint as in Model \ref{model:mean-shift}, which renders most changepoint algorithms based on such an assumption ineffective.}
    \label{fig-realdatawell34}
\end{figure}

We obtain 38 functional connectivity networks on 181 vertices across 246 days. Since the alignment of vertices among the networks is unknown, we apply the Fast Approximate Quadratic (FAQ) algorithm \cite{Vogelstein2015-cx} to perform graph matching on all networks. We avoid the birth and death period, and choose only $m=30$ graphs. We find the largest common connected component among the aligned networks, which contains $n=44$ nodes, and we compute the estimated iso-mirror following Algorithm \ref{alg:1}. Finally, we fit the data with $l_{\infty}$ regression, specifying only 1 changepoint following Algorithm \ref{alg:2}. This changepoint is estimated to occur on day 156 (see Figure \ref{fig-realdatawell34}). On the right panel, we also plot the control chart, which is simply the Frobenius norm difference between the adjacency matrices of two consecutive time points. Notably, this Frobenius norm difference consistently deviates from zero across almost all pairs of consecutive points. This pattern indicates that the time series of functional connectivity networks is in a state of constant flux. As a result, the zeroth-order changepoint Model \ref{model:mean-shift}, which would effectively classify every point as a changepoint, is not appropriate. Additional features and summary statistics of this time series of organoid networks are provided in Appendix~\ref{sec:organoiddetails}.

\section{Conclusion and Discussion}
\label{Sec:Con_and_dis}

Changepoint localization for network time series is complicated by the fact that a changepoint can be more delicate than a simple shift in network distribution. For latent position process networks, we address three core points in this work: we formalize the concept of changepoints of different orders; we construct, in Model \ref{changepointmodel}, a time series of latent position networks with a first-order changepoint; and we show that such a changepoint can be successfully localized through spectral decompositions of network observations and spectrally-derived estimates of the associated Euclidean mirror. Building on work in \cite{athreya2025euclidean}, we also define asymptotic Euclidean realizability for a network time series, highlighting the significance of both network size $n$ and the number of sampled time points $m$ in such a time series.

The Euclidean mirror uses classical multidimensional scaling to represent a dissimilarity measure for latent position networks, and thus depends on the choice of embedding dimension and network dissimilarity. In our simulations, we underscore the efficacy of our mirror-based changepoint localization, with covariance dissimilarity, for Model \ref{changepointmodel} and also illustrate the bias-variance trade-off associated to the embedding dimension. In our real data analysis of brain organoid networks, the mirror exhibits an approximately piecewise linear structure, and our algorithm locates a changepoint. Obtaining similarly powerful results for other kinds of data will require detailed analysis of different network dissimilarities and embedding dimensions. 

There are multiple axes along which these results can be generalized and improved. In particular, finding a sharper bound in Theorem \ref{thm:consistency asy1-d} requires improved distributional results on $\hat{\psi}-\psi$. Secondly, improved understanding of the effects of ISOMAP on the mirror estimate, particularly its impact on aggregating errors between the estimated and true mirror, will lead to improved understanding of the iso-mirror as a tool for network time series analysis. 

The current paper focuses on the task of changepoint localization, but the problem of changepoint detection is also important, and would benefit from further investigation. A related issue on this point is isolating and locating an indeterminate number of changepoints, which requires refined estimates beyond those we consider here. The success of mirror-based estimation for simple latent position networks and the associated provable guarantees for localization suggest that these techniques have wider promise in the analysis of evolving networks.

\begin{appendix}\label{sec:Appendix}
\section{Proofs of some basic lemmas}
Proof of Lemma~\ref{lem:deterministic_LPP}.
\begin{proof}
When $\varphi(t)$ is deterministic, for any $x,y$:
\begin{align*}
d_{MV}\left(\varphi(x),\varphi(y)\right)^2 &=\min_{W \in \mathcal{O}^{d \times d}} \left\Vert \EE[(\varphi(x)-W\varphi(y) )(\varphi(x)-W \varphi(y) )^{\top}]\ \right\Vert_2  \\
& = \min_{W \in \mathcal{O}^{d \times d}} \left\Vert \left(\varphi(x)-W\varphi(y) \right) \left(\varphi(x)-W\varphi(y) \right)^{\top} \right\Vert _2  \\
& =  \min_{W \in \mathcal{O}^{d \times d}} \|\varphi(x)-W\varphi(y)\|^2_2 \\
&= (\| \varphi(x) \|_2- \| \varphi(y) \|_2)^2.
\end{align*}
The last equality follows since $ \|\varphi(x)-W\varphi(y)\|_2^2  \geq (\| \varphi(x) \|_2- \| \varphi(y) \|_2)^2  $ always holds and there exists a $W$ such that $\varphi(x)$ and $\varphi(y)$ are linearly dependent, in which case the lower bound is achieved. The result now follows after centering the mirror.
\end{proof}
Proof of Lemma~\ref{lem:dmv-nochangepoint}.
\begin{proof}
Suppose $t,t' \in \{1,2,\cdots,n\}$ and $t'<t$. Put $a:=t-t'$. Since $X_t$ is a one-dimensional random variable,
\begin{align*}
d_{MV}\left(X_t,X_{t'}\right)^2 &=\min_{W \in \mathcal{O}^{1 \times 1}} \left\Vert\EE[\left(X_t-W X_{t'}\right)(X_t-W X_{t'})^{\top}] \right\Vert_2  \\ 
&=\min_{W} \left|\EE[X_t^2-2WX_tX_{t'}+W^2X_{t'}^2]\right|.
\end{align*}

To address the minimization over orthogonal matrices, observe that in one dimension, orthogonal transformations correspond only to multiplication by $1$ or $-1$. So we need only consider the two cases $W=1$ or $W=-1$. Since $X_t \geq 0$ for all $t$, $\EE[X_tX_{t'}]\geq 0$. Thus $W=1$ minimizes the above, and we conclude that 
$$
d_{MV}(X_t,X_{t'})^2=\EE[(X_t-X_{t'})^2].
$$

The increment $X_t-X_{t'}$ satisfies $X_t-X_{t'}=V_a\delta$, where $V_a$ is a Binomial random variable with $a$ trials and success probability $p$, namely $V_a \sim \text{Binomial}(a,p)$. Thus
\begin{align*}
\EE[(X_t-X_{t'})^2]&=
\delta^2\EE[V_a^2] \\
&=\left(a^2p^2+a\left(p-p^2\right)\right)\delta^2.
\end{align*}

\end{proof}

\begin{lemma}\label{lem:deterministic_LPP}
Consider $t_{\beta,m} = \lfloor m\beta \rfloor$ with $ \beta \in [0,1]$ fixed as $m \to \infty$. Denote by $X^{(m)}_{t_{\beta,m}}$ be the latent position process for Model \ref{no-changepoint}, for a given $m$, evaluated at $t_{\beta,m}$. Let $\{Z_t, 0 \leq t \leq 1\}$ denote the continuous deterministic LPP defined by $Z_t=tp$ . Then $X^{(m)}_{t_{\beta,m}} \overset{L^2}{\longrightarrow} Z_\beta$ for any $\beta$ as $m \to \infty$. 
Next, Consider $t_{\beta,m} = \lfloor m\beta \rfloor$ with $ \beta \in [0,1]$ fixed and $t^*_m= \lfloor m t^* \rfloor$ with $t^* \in [0,1]$ as $m \to \infty$. Suppose $\tilde{X}^{(m)}_{t_{\beta, m}}$ is the latent position process for Model \ref{changepointmodel} with change of jump probability at $t^*_m$, for a given $m$ at $t_{\beta,m}$. Define the deterministic latent position process by 
$$
Z_{\beta} = 
\begin{cases}
    p\beta \mbox{~for~} \beta \leq t^* \\
    q\beta+(p-q)t^* \mbox{~for~} \beta > t^*,
\end{cases}
$$ where $p$, $q$ are as in Model \ref{changepointmodel}. Then for each $\beta$, $\tilde{X}^{(m)}_{t_{\beta, m}} \overset{L^2}{\longrightarrow} Z_\beta$ as $m \to \infty$. 
\end{lemma}
\begin{proof}
First we consider Model \ref{no-changepoint}. Fix $\beta \in [0,1]$. Put $t_{\beta,m} = \lfloor m\beta \rfloor $. Then $X^{(m)}_{t_{\beta,m} }=\frac{V_{t_{\beta,m},p}}{m}$, where $V_{t_{\beta,m}}$ is a Binomial random variable with $t_{\beta,m}$ trials and success probability $p$. Therefore, $\EE[X^{(m)}_{t_{\beta,m} }]=\frac{\lfloor m \beta \rfloor p }{m}$ and as $m \to \infty$,
\begin{align*}
\EE[X^{(m)}_{t_{\beta,m}}-\beta p]^2&= \EE[X^{(m)}_{t_{\beta,m}}-\EE[X^{(m)}_{t_{\beta,m}}] + \EE[X^{(m)}_{t_{\beta,m}}] - \beta p ]^2\\
&= \frac{ \lfloor m \beta \rfloor p (1-p)  }{m^2} + \left( \frac{\lfloor m \beta \rfloor p }{m}-
\beta p \right)^2 \to 0.
\end{align*}
Thus $ X^{(m)}_{t_{\beta,m}} \overset{L^2}{\longrightarrow} \beta p=Z_\beta$. 

Now consider Model \ref{changepointmodel}. Again fix $t^*, \beta \in [0,1]$ and put $t_{\beta,m}=\lfloor m\beta \rfloor$ and $t^*_m=\lfloor mt^* \rfloor$. If $\beta \leq t^*$, then $t_{\beta,m}=\lfloor\beta m \rfloor \leq \lfloor t^* m \rfloor = t^*_m$; thus $\tilde{X}^{(m)}_{t_{\beta, m}}=X^{(m)}_{t_{\beta, m}}$ and the result follows from what we have proved above.

When $\beta > t^*$, note for any $x \in \R$, $\lfloor x \rfloor -1 < x \leq \lfloor x \rfloor$, and thus $\lfloor\beta m \rfloor -\lfloor t^* m \rfloor > m\beta-(mt^*+1)$. Then for all $m> \frac{1}{\beta - t^*}$, $t_{\beta,m} > t^*_m$. In this case, $\tilde{X}^{(m)}_{t_{\beta, m}} = \frac{V_{t^*_m,p}+ V_{t^*_m-t_{\beta,m},q}  }{m}$. Therefore $\EE[\tilde{X}^{(m)}_{t_{\beta, m}}]= \frac{\lfloor mt^* \rfloor p + ( \lfloor m\beta \rfloor -\lfloor mt^*  \rfloor )q }{m}$ and as $m \to \infty$,
$$
\EE[\tilde{X}^{(m)}_{t_{\beta,m}}-(t^*p+\beta q - t^*q)]^2 = \EE \left[ \left[\frac{V_{t^*_m,p}}{m} - t^*p\right] + \left[ \frac{V_{t^*_m-t_{\beta,m},q}}{m} -(\beta q - t^*q) \right] \right]^2.
$$
Both squared terms $\EE\left[\frac{V_{t^*_m,p}}{m} - t^*p\right]^2$ and $\EE\left[ \frac{V_{t^*_m-t_{\beta,m},q}}{m} -(\beta q - t^*q) \right]^2$ converge to zero by the same argument as before. For the cross term,\\ $\EE \left[\frac{V_{t^*_m,p}}{m} - t^*p\right] \left[ \frac{V_{t^*_m-t_{\beta,m},q}}{m} -(\beta q - t^*q) \right]$, note that $V_{t^*_m,p}$ is independent of \\$V_{t^*_m-t_{\beta,m},q}$, so the expectation of the product is equal to the product of the expectation, with converging to zero.

\end{proof}

\section{Proof of Theorem \ref{thm:ourmodelisasysum}}\label{sec:proof_of_thm2}
\begin{proof}
We first state certain properties of the two distance matrices $\mathcal{D}_{\varphi_m}$ and  $\mathcal{D}_Z$. Denote the eigenvalues and the corresponding orthonormal eigenvectors of $-\frac{1}{2}P\mathcal{D}^{(2)}_ZP$ and $-\frac{1}{2}P\mathcal{D}^{(2)}_{\varphi_m}P$ respectively by $\lambda_1, \lambda_2, \cdots, \lambda_m$ and $U_1,U_2,\cdots,U_m$; $\tilde{\lambda}_1, \tilde{\lambda}_2,\cdots, \tilde{\lambda}_m$ and $\tilde{U}_1,\tilde{U}_2,\cdots,\tilde{U}_m$. Then the $k$th dimension of the mirror for Model \ref{infillmodel} is: $\sqrt{\tilde{\lambda}_k}\tilde{U}_k $. Since $\mathcal{D}_Z$ is exactly Euclidean 1-realizable, by Theorem \ref{thm:PSD-iffrealizable}, its mirror is $ \sqrt{\lambda_1}U_1$. 

\begin{property}\label{Prop:1}
$\lambda_1 \sim \Theta(m)$, $\lambda_i =0$ for $i \ge 2$. And $ \left\Vert U_1 \right\Vert_{2 \to \infty} \sim O\left(\frac{1}{\sqrt{m}}\right)$. 
\end{property}

\begin{proof}
Since $\mathcal{D}_Z$ is exactly 1-d Euclidean realizable with mirror $\sqrt{\lambda_1}U_1$,\\ $-\frac{1}{2}P\mathcal{D}^{(2)}_ZP=\lambda_1U_1U_1^{\top}$ where $\sqrt{\lambda_1}U_1=\left(\psi_Z(\frac{1}{m}),\psi_Z(\frac{2}{m}),...\psi_Z(\frac{m}{m})\right)^{\top}$. Thus, since $\psi_Z^2$ is Riemann integrable,
$$
\frac{\lambda_1}{m}= \frac{\sum^m_{i=1}\psi_Z^2\left(\frac{i}{m}\right)}{m} \to \int^1_0\psi_Z^2(x) dx =C \text{~ as ~} m \to \infty.
$$ That is, $\lambda_1 \sim \Theta(m)$. Next, since \begin{align*}
\left\Vert\sqrt{\lambda_1}U_1\right\Vert_{2 \to \infty} &= \left\Vert\left(\psi_Z\left(\frac{1}{m}\right),\psi_Z\left(\frac{2}{m}\right),...,  \psi_Z\left(\frac{m}{m}\right)\right)^{\top}\right\Vert_{2 \to \infty}\\
&\leq \max_{t \in [0,1]} |\psi_Z(t)|  \leq p+q+c_0,
\end{align*} thus $\left\Vert U_1 \right\Vert_{2 \to \infty} \sim O\left(\frac{1}{\sqrt{m}}\right)$.
\end{proof}

\begin{property}\label{Prop:2}
$\|\mathcal{D}^{(2)}_{\varphi_m}-\mathcal{D}^{(2)}_Z\|_{\infty} \sim O(1)$, and $\tilde{\lambda}_1 \sim \Theta(m)$,  $\tilde{\lambda}_i \sim O(1)$ for $i \ge 2$.
\end{property}
\begin{proof}

Because $\left(\mathcal{D}^{(2)}_{\varphi_m}-\mathcal{D}^{(2)}_Z\right)_{i,j} \sim O\left(\frac{1}{m}\right)$ for all $i,j$, we get $\left\Vert\mathcal{D}^{(2)}_{\varphi_m}-\mathcal{D}^{(2)}_Z\right\Vert_{\infty} \sim O(1)$ and $\left\Vert\mathcal{D}^{(2)}_{\varphi_m}-\mathcal{D}^{(2)}_Z\right\Vert_{F} \sim O(1)$.
Also note $\|P\|_2=1$, so $$\adjustbox{width=4.513in}{$\displaystyle \left\Vert\frac{1}{2}P\left(\mathcal{D}^{(2)}_{\varphi_m}-\mathcal{D}^{(2)}_Z\right)P\right\Vert^2_F \leq \left\Vert\frac{1}{2}\left(\mathcal{D}^{(2)}_{\varphi_m}-\mathcal{D}^{(2)}_Z\right)\right\Vert^2_F\|P\|^4_2= \left\Vert\frac{1}{2}\left(\mathcal{D}^{(2)}_{\varphi_m}-\mathcal{D}^{(2)}_Z\right)\right\Vert^2_F \sim O(1).$}$$ Thus 
\begin{align*}
\left\Vert\frac{1}{2}P(\mathcal{D}^{(2)}_{\varphi_m}-\mathcal{D}^{(2)}_Z)P\right\Vert^2_2 \leq \left\Vert\frac{1}{2}P(\mathcal{D}^{(2)}_{\varphi_m}-\mathcal{D}^{(2)}_Z)P\right\Vert^2_F \sim O(1).
\end{align*}

Weyl's theorem guarantees that $\left|\lambda_k-\tilde{\lambda}_k\right|\leq \left\Vert\frac{1}{2}P(\mathcal{D}^{(2)}_{\varphi_m}-\mathcal{D}^{(2)}_Z)P\right\Vert_2$ for all $k$. Since $\lambda_1 \sim \Theta(m)$ and  $\lambda_i =0$ for $i \ge 2$, the result follows.
\end{proof}

\begin{property}\label{Prop:3}
There exists a $w=1$ or $w=-1$ such that:\\$\| \sqrt{\tilde{\lambda}_1}\tilde{U}_1-w\sqrt{\lambda_1}U_1 \|_{2\to\infty} \to 0 $ as $m \to \infty$.
 \end{property}

\begin{proof}
From Property \ref{Prop:1}, we know 
$\lambda_1 \sim \Theta(m)$ and $\left\Vert U_1 \right\Vert_{2 \to \infty} \sim O(\frac{1}{\sqrt{m}})$. There exists a $w=1$ or $w=-1$ such that:
\begin{align*}
&\left\Vert \sqrt{\tilde{\lambda}_1}\tilde{U}_1-w\sqrt{\lambda_1}U_1  \right\Vert_{2 \to \infty}  \leq \sqrt{ \tilde{\lambda}_1} \left\Vert \tilde{U}_1-wU_1\right\Vert_{2 \to \infty} + \left|\sqrt{\tilde{\lambda}_1} -\sqrt{\lambda_1}\right|\left\Vert U_1 \right\Vert_{2 \to \infty}  \\
&\leq  \sqrt{ \tilde{\lambda}_1}\frac{c\left\Vert\frac{1}{2}P(\mathcal{D}^{(2)}_{\varphi_m}-\mathcal{D}^{(2)}_Z)P\right\Vert_{\infty} \left\Vert U_1 \right\Vert_{2 \to \infty} }{\lambda_1}
+\left|\sqrt{\tilde{\lambda}_1} -\sqrt{\lambda_1}\right|\left\Vert U_1 \right\Vert_{2 \to \infty} \\
& \leq \left( \frac{ c \sqrt{ \tilde{\lambda}_1} \left\Vert\frac{1}{2}P(\mathcal{D}^{(2)}_{\varphi_m}-\mathcal{D}^{(2)}_Z)P\right\Vert_{\infty}}{\lambda_1}+\left|\sqrt{\tilde{\lambda}_1} -\sqrt{\lambda_1}\right| \right) \left\Vert U_1 \right\Vert_{2 \to \infty}. 
\end{align*}
The second inequality uses Theorem 4.2 in \cite{cape2019two}, which we quote here: 
\begin{theorem}
Let \( X, E \in \mathbb{R}^{p \times p} \) be symmetric matrices where \(\text{rank}(X) = r\) and \(X\) has spectral decomposition $X = U \Lambda U^T$ and $X+E=\hat{U} \hat{\Lambda} \hat{U}^T+\hat{U}_{\perp} \hat{\Lambda}_{\perp} \hat{U}_{\perp}^T$. Suppose the leading eigenvalues of $X$ are given by \( |\lambda_1| \geq |\lambda_2| \geq \ldots \geq |\lambda_r| \geq 0 \). If \( |\lambda_r| \geq 4 \|E\|_{\infty} \), then there exists an orthogonal matrix \( W_U \in O_r \) such that
$$
    \left\Vert \hat{U} - U W_U \right\Vert_{2 \to \infty} \leq 14 \left( \frac{\|E\|_{\infty}}{\left|\lambda_r\right|} \right) \|U\|_{2 \to \infty}.
$$
\end{theorem}

The condition that 
$|\lambda_1|>4\left\Vert P\left(\mathcal{D}^{(2)}_{\varphi_m}-\mathcal{D}^{(2)}_Z\right)P\right\Vert_{\infty}$ is satisfied since $\lambda_1 \sim \Theta(m)$ and $$\left\Vert P\left(\mathcal{D}^{(2)}_{\varphi_m}-\mathcal{D}^{(2)}_Z\right)P\right\Vert_{\infty} \leq \left\Vert\mathcal{D}^{(2)}_{\varphi_m}-\mathcal{D}^{(2)}_Z\right\Vert_{\infty}\left\Vert P\right\Vert^2_{\infty} \leq 4\left\Vert\mathcal{D}^{(2)}_{\varphi_m}-\mathcal{D}^{(2)}_Z\right\Vert_{\infty} \sim O(1).$$

Now we consider $\left|\sqrt{\tilde{\lambda}_1} -\sqrt{\lambda_1}\right|$. Note that $$\left|\sqrt{\tilde{\lambda}_1} -\sqrt{\lambda_1}\right| = \frac{\left|\tilde{\lambda}_1-\lambda_1\right|}{\sqrt{\tilde{\lambda}_1}+\sqrt{\lambda_1}} \sim \frac{O(1)}{\Theta\left(\sqrt{m}\right)}  \sim O\left(\frac{1}{\sqrt{m}}\right).$$  Thus for large $m$, there exists a constant $C$ such that $ \left| \sqrt{\tilde{\lambda}_1}-\sqrt{\lambda}_1 \right| \leq \frac{C}{\sqrt{m}}$, and the above term can be further bounded by
$$
\adjustbox{width=4.513in}{$\displaystyle
\left( \frac{ c\left(\sqrt{\lambda_1}+\frac{C}{\sqrt{m}} \right) \left\Vert \frac{1}{2}P(\mathcal{D}^{(2)}_{\varphi_m}-\mathcal{D}^{(2)}_Z)P \right\Vert _{\infty}}{\lambda_1}+\left|\sqrt{\tilde{\lambda}_1} -\sqrt{\lambda_1}\right| \right) \left\Vert U_1 \right\Vert_{2 \to \infty} \sim O\left(\frac{1}{m}\right)$}
$$ 
as required. 
\end{proof}

\begin{property}\label{Prop:4}
 $\frac{ \sum^m_{i=2} \tilde{\lambda}^2_i  }{ \tilde{\lambda}_1 } \to 0$ as $m \to \infty$.
 \end{property}
\begin{proof}
By the Hoffman-Wielandt inequality \cite{HJ85_matrix}, we derive 
$$
\sum^m_{i=2} \left(\tilde{\lambda}_i- \lambda_i\right)^2 \leq \left\Vert\frac{1}{2}P\left(\mathcal{D}^{(2)}_Z-\mathcal{D}^{(2)}_{\varphi_m}\right)P \right\Vert ^2_F \sim O(1).
$$
Note $\lambda_i=0$ for $i \geq 2$. Thus we have 
$$
\sum^m_{i=2} \tilde{\lambda}_i^2 \sim O(1).
$$
Recall from Property \ref{Prop:2} that $\tilde{\lambda}_1 \sim \Theta(m)$; then  $\frac{ \sum^m_{i=2} \tilde{\lambda}^2_i  }{ \tilde{\lambda}_1 } \sim O\left(\frac{1}{m}\right)$ and the result follows. 
\end{proof}

Now we prove each of the claims in Theorem \ref{thm:ourmodelisasysum}. 
Note that Claim (2) is immediate because $$\max_{i,j\in \{1,2,\cdots,m\}} 
 \left|\left(\mathcal{D}^{(2)}_{\varphi_m}-\mathcal{D}_Z^{(2)}\right)_{i,j}\right| \leq \frac{1}{4m}$$ for any $m$. Claims (3) and (4) are proved in Property \ref{Prop:2} and Property \ref{Prop:4}.

For Claim (1), recall $\tpsim_1$ is the piecewise linear interpolation between the points $$\left\{ \left(t_i, \left( \sqrt{\tilde{\lambda}_1}\tilde{U}_1 \right)_i\right), 1 \leq i \leq m\right
\},$$
Consider the unique extension of this function to a function that is linear on interval $[0, t_2]$; through this extension, we define $\tpsim_1$ on $[0,t_1]$ on as well. 
Denote by $\tilde{\psi}_{Z}$ the piecewise linear interpolation between the points\\ $\left\{\left(t_i, \left( \sqrt{\lambda_1} U_1 \right)_i\right), 1 \leq i \leq m\right\}=\left\{\left(t_i, \psi_Z\left(\frac{i}{m}\right)  \right), 1 \leq i \leq m\right\}$. Then 
$$
\sup_{t \in [0,1]} \left|\tpsim_1(t)-\psi_Z(t)\right| \leq \sup_{t \in [0,1]}\left|\tpsim_1(t)-\tilde{\psi}_Z(t)\right| + \sup_{t \in [0,1]}\left|\tilde{\psi}_Z(t)-\psi_Z(t)\right|. 
$$
Observe that $\tpsim_1(t)$ and $\tilde{\psi}_Z$ are both piecewise linear interpolations at the same time points $\{t_1,t_2,\cdots,t_m\}$, where $t_m=1$. By inearity on $[0,t_2]$, we have $$\left|\tpsim_1(0)-\tilde{\psi}_Z(0)\right| \leq 3\max_{t\in\{t_1,t_2\}}\left|\tpsim_1(t)-\tilde{\psi}_Z(t)\right|.$$ 
Now because of Lemma \ref{Lem:endpointlemma} and Property \ref{Prop:3}, we see that

\begin{align*}
\sup_{t \in [0,1]}\left|\tpsim_1(t)-\tilde{\psi}_Z(t)\right| &= \max_{t\in\{0,t_1,t_2,\cdots,t_m\}} \left|\tpsim_1(t)-\tilde{\psi}_Z(t)\right|\\
&\leq 3 \left\Vert \sqrt{\tilde{\lambda}_1}\tilde{U}_1-\sqrt{\lambda_1}U_1 \right\Vert_{2\to\infty} \to 0.
\end{align*}

Consider $\sup_{t \in [0,1]}\left|\tilde{\psi}_Z(t)-\psi_Z(t)\right|$. Recall that $\tilde{\psi}_Z(t)=\psi_Z(t)$ for \\ $t\in\{t_1,t_2,\cdots,t_m\}$ and both of them are piecewise linear. 

Then $\sup_{t \in [0,1]}|\tilde{\psi}_Z(t)-\psi_Z(t)|>0$ only if $t^* \notin \{t_1,t_2,\cdots,t_m\}$. When $t^* \notin \{t_1,t_2,\cdots,t_m\}$,
$\tilde{\psi}_Z$ and $\psi_Z$ disagree only on the interval $[t_i,t_{i+1}]$ that contains $t^*$. We can bound the difference by 
$$
\sup_{t \in [0,1]}|\tilde{\psi}_Z(t)-\psi_Z(t)|=\biggl|\frac{(p-q)(t_{i+1}-t^*)(t^*-t_i)}{t_{i+1}-t_i}\biggr| \leq |p-q||t_{i+1}-t_i|=\frac{|p-q|}{m}, 
$$
because $t_i=\frac{i}{m}$ for all $i \in \{1,\cdots,m\}$. Thus this term also tends to 0 as $m \to \infty$, as required.

\end{proof}

\section{Proof of Corollary \ref{cor:simple_upperbound}}
\begin{proof}
We first find an upper bound for $B$: note that $$ \min_{R\in \mathcal{O}^{c \times c}} \|\hat{U}-UR\|_F \leq \|\hat{U}-U\|_F \leq  \|\hat{U}\|_F + \|U\|_F \leq 2\sqrt{c}$$ because $\hat{U}$ and $U$ both have 
$c$ orthonormal columns. Thus $B\leq\sqrt{2c}$. Since $c=1$, $B\leq\sqrt{2}$ and $\kappa=1$. Additionally, we have  $\lambda_1\left(E_{\varphi}\right) \sim \Theta(m)$, and since $\frac{\log(n)}{\sqrt{n}} \leq 8$ for $n \in \mathbb{N}^+$, we get
\begin{align*}
\sum_{i=1}^{m} \|\hat{\psi}(t_i)-R\psi(t_i)\|^2 
& \leq B^2\lambda_1(E_{\varphi})\left(2+4B\kappa^{1/2}+(1+2B)\frac{m\log(n)}{\sqrt{n}\lambda_1(E_{\varphi})}\right)^2 \\ 
& \leq B^2 \lambda_1(E_{\varphi}) \left(2+4\sqrt{2}+\left(1+2\sqrt{2}\right)8 \right)\\
& \leq C B^2 \lambda_1(E_{\varphi}).
\end{align*}
Then plug $B=\frac{2^{3/2}}{\lambda_1(E_{\varphi})}\left(\frac{m\log(n)}{\sqrt{n}}+\left(\sum_{i=c+1}^{m}\lambda^2_i\left(E_{\varphi}\right)\right)^{1/2}\right)$ back we get the result.
\end{proof}

\section{Supporting lemmas for Section~\ref{Sec:consistency}}
\begin{lemma} \label{Lem:endpointlemma}
    Consider two continuous piecewise linear functions $f_1$ and $f_2$ defined on $x \in [a,b]$, and each with a finite collection of changepoint sets denoted, respectively, by $\{t_1,t_2,\cdots,t_m\}=\mathcal{T}_1$ and  $\{t'_1,t'_2,\cdots,t'_n\}=\mathcal{T}_2$. Suppose $\mathcal{T}=\mathcal{T}_1 \cup \mathcal{T}_2$. Then 
    $$
    \sup_{x \in [a,b]} |f_1(x)-f_2(x)| = \max\{ |f_1(a)-f_2(a)|,\max_{x \in \mathcal{T}} |f_1(x)-f_2(x)|, |f_1(b)-f_2(b)|\},
    $$
    namely, the uniform norm of difference of two piecewise linear functions on an interval are determined only by their changepoints and endpoints of the interval.  
\end{lemma}
\begin{proof}
Order the set $\mathcal{T}$ as $\mathcal{T}=\{x_1,x_2,\cdots,x_k\}$, and $x_i \leq x_{i+1}$ for all $1\leq i \leq k-1 $. Note $k \leq m+n$. Then on any $[x_i,x_{i+1}]$, $f_1$ and $f_2$ both are linear functions; hence the function $f_1(x)-f_2(x)$ is also a linear function on $[x_i,x_{i+1}]$. Further, the function $|f_1(x)-f_2(x)|$ is convex on the compact and convex set $[x_i,x_{i+1}]$. Thus its maximum is obtained at some extreme point of the set $[x_1,x_{i+1}]$; these extreme points are $\{x_i,x_{i+1}\}$. We have that $$
\sup_{x \in [x_i,x_{i+1}]}|f_1(x)-f_2(x)|=\max\{ |f_1(x_i)-f_2(x_i)|, |f_1(x_{i+1})-f_2(x_{i+1})|\}.
$$

This also holds on $[a,x_1]$ and $[x_k,b]$. Partitioning the interval $[a,b]$ as $[a,b]=[a,x_1] \cup [x_1,x_2] \cdots \cup [x_k,b] $, we derive  
\begin{align*}
&\sup_{x \in [a,b]}|f_1(x)-f_2(x)| \\
&=    \max \{ \sup_{t \in [a,x_1]}|f_1(x)-f_2(x)| , \sup_{t \in [x_1,x_2]}|f_1(x)-f_2(x)|, \cdots \sup_{t \in [x_k,b]}|f_1(x)-f_2(x)|\}  \\
&=  \max\{ |f_1(a)-f_2(a)|,\max_{x \in \mathcal{T}} |f_1(x)-f_2(x)|, |f_1(b)-f_2(b)|\}. 
\end{align*}
\end{proof}
\begin{lemma}\label{lem:consistency}
Let $\psi(t)=at+b+\Delta(t-t^*)I_{\{t>t^*\}},$ $\Delta\neq0$, and let $\Tilde{\hat{\psi}}$ be the continuous piecewise linear interpolation of $\left\{\left(t_i, \hat{\psi}(t_i)\right)| 1 \leq i \leq m\right\}$. Set $$ \rho=\mathop{max}\limits_{1\leq i \leq m-1} |t_i-t_{i+1}|.$$ Then the $\ell_{\infty}$ changepoint localization estimator $\hat{t}$ satisfies 
$$
\left|\hat{t}-t^*\right| \leq  \left( \frac{4 \min_{w \in \{\pm1\}  } \sup_{t \in [0,T]} \left| \Tilde{\hat{{\psi}}}(t) - w\psi(t)\right|}{|\Delta|} + 2\rho \right) \frac{T}{ min\{T-t^*,t^*\} }.
$$   
\end{lemma}
\begin{proof}
Without loss of generality, assume $0<\hat{t}< t^*$.  Next, because $h \in \mathcal{S}(\{\hat{t}\}) $, we have:  
$$
\min_{f \in \mathcal{S}(\{\hat{t}\})} \sup_{t \in [0,T]} |f(t)-\psi(t)| \leq \sup_{t \in [0,T]}|h(t)-\psi(t)|.
$$ 
The minimization on the left hand side may be written as:
\begin{align*}
&\min_{f \in \mathcal{S}(\{\hat{t}\})} \sup_{t \in [0,T]} |f(t)-\psi(t)| \\
&\adjustbox{width=4.513in}{$\displaystyle = \min_{\alpha,\beta_L,\beta_R,\beta_L \neq \beta_R }\sup_{t \in [0,T]} |\alpha+\beta_L (t-\hat{t})+(\beta_R-\beta_L)(t-\hat{t})I_{\{t>\hat{t}\}}-at-b-\Delta(t-t^*)I_{\{t>t^*\} }|$}\\
&\adjustbox{width=4.513in}{$\displaystyle = \min_{\alpha,\beta_L,\beta_R,\beta_L \neq \beta_R }\sup_{t \in [0,T]} |\alpha'+\beta_L' (t-\hat{t})+(\beta_R'-\beta_L')(t-\hat{t})I_{\{t>\hat{t}\}}-|\Delta|(t-t^*)I_{\{t>t^*\} }|,$}%
\end{align*}
where we have defined $\alpha'=\mathrm{sgn}(\Delta)(\alpha-b)$, $\beta_L'=\mathrm{sgn}(\Delta)(\beta_L-a),$ $\beta_R'=\mathrm{sgn}(\Delta)(\beta_R-a).$ Applying Lemma \ref{Lem:endpointlemma}, we see that this is equivalent to 
\begin{align*}
&\min_{f \in \mathcal{S}(\{\hat{t}\})} \sup_{t \in [0,T]} |f(t)-\psi(t)| \\
&\adjustbox{width=4.513in}{$\displaystyle = \min_{\alpha',\beta_L',\beta_R', \beta_L' \neq \beta_R'} \max \{ |\alpha'|,|\alpha'-\beta_L'\hat{t}|, |\alpha'+\beta_R'(t^*-\hat{t})|, \left|\alpha'+\beta_R'(T-\hat{t})-|\Delta|(T-t^*) \right| \}.$}
\end{align*} By checking the Karush-Kuhn-Tucker (KKT) conditions (see Appendix \ref{Sec:append-KKT}), we find that the minimum is given by
$$\frac{T-t^*}{2(T-\hat{t})} |\hat{t}-t^*||\Delta|$$ 
from which we conclude that
$$
\frac{T-t^*}{2(T-\hat{t})} |\hat{t}-t^*||\Delta|   \leq \sup_{t \in [0,T]} |h(t)-\psi(t)|.
$$
To bound the right hand side, we introduce $\psi^T(t)$ as a translation of $\psi(t)$ such that the changepoint is at the sampled time $t_{k^*}$ that is closest to $t^*$, the changepoint of the true $\psi(t)$. That is: 
$$
\psi^T(t)= at+b+\Delta(t-t_{k^*})I_{\{t>t_{k^*}\}}.
$$
Note $\psi^T(t) \in \mathcal{S}(\mathcal{T})$ and recall $\max_{t \in \mathcal{T} } | h - \hat{\psi}(t)| = \min_{f \in \mathcal{S}(\mathcal{T})}  \max_{t \in \mathcal{T} } | f - \hat{\psi}(t)|$.
From this, we find that
$$
\max_{t \in \mathcal{T} } | h(t)-\hat{\psi}(t) | \leq \max_{t \in \mathcal{T} }   |\psi^T(t)-\hat{\psi}(t)|.
$$
We also have $\sup_{t \in [0,T]} |\psi^T(t)-\psi(t)|=|\Delta| |t^*-t_{k^*}| \leq|\Delta| \rho $.

Because  $h(t)-\Tilde{\hat{\psi}}(t)$ and $\psi^T(t)-\Tilde{\hat{\psi}}(t)$ are both continuous piecewise linear functions with changepoints in $\mathcal{T}$, again Lemma \ref{Lem:endpointlemma} implies that $$
\max_{t \in \mathcal{T} } | h(t)-\hat{\psi}(t) | = \sup_{t \in [0,T]} | h(t)-\Tilde{\hat{\psi}}(t) |,\;
\max_{t \in \mathcal{T} }   |\psi^T(t)-\hat{\psi}(t) | = \sup_{t \in [0,T]} | \psi^T(t)-\Tilde{\hat{\psi}}(t) |. 
$$ Therefore, \begin{equation} \label{dumb_bound}
\sup_{t \in [0,T]} | h(t)-\Tilde{\hat{\psi}}(t) | \leq \sup_{t \in [0,T]} | \psi^T(t)-\Tilde{\hat{\psi}}(t)|.     
\end{equation} Next, note that 
\begin{align}
\frac{T-t^*}{2(T-\hat{t})} |\hat{t}-t^*| |\Delta|  
&\leq  \sup_{t \in [0,T]} |h(t)-\psi(t)|  \notag \\
&\leq  \sup_{t \in [0,T]} |h(t)-\Tilde{\hat{\psi}}(t)| + \sup_{t \in [0,T]}|\Tilde{\hat{\psi}}(t)-\psi(t)| \notag \\
&\leq \sup_{t \in [0,T]} | \psi^T(t)-\Tilde{\hat{\psi}}(t) |  + \sup_{t \in [0,T]} |\Tilde{\hat{\psi}}(t)-\psi(t)| ~\text{by Eq } (\ref{dumb_bound}) \notag \\
&\adjustbox{width=3.5in}{$\displaystyle \leq \sup_{t \in [0,T]} | \psi^T(t)-\psi(t)|+\sup_{t \in [0,T]} |\psi(t)-\Tilde{\hat{\psi}}(t) |  + \sup_{t \in [0,T]} |\Tilde{\hat{\psi}}(t)-\psi(t)| $} \notag \\
&= \sup_{t \in [0,T]} |\psi^T(t)-\psi(t)|+2\sup_{t \in [0,T]} |\Tilde{\hat{\psi}}(t)-\psi(t)| \notag \\
&\leq |\Delta| \rho  +2\sup_{t \in [0,T]} |\Tilde{\hat{\psi}}(t)-\psi(t)|. \notag
\end{align} Dividing both sides by $\frac{T-t^*}{2(T-\hat{t})} |\Delta|$ and recalling that 
$$\frac{T-\hat{t}}{T-t^*} \leq \frac{T}{ min\{t^*,T-t^*\}},$$ we conclude that $$
|\hat{t}-t^*| \leq  \left( \frac{4 \sup_{t \in [0,T]} | \Tilde{\hat{{\psi}}}(t) - \psi(t)|}{|\Delta|} + 2\rho \right) \frac{T}{ min\{T-t^*,t^*\} }.
$$
For any given $\{\hat{\psi}(t_i) | 1 \leq i \leq m\}$ we can also use $\{-\hat{\psi}(t_i) | 1 \leq i \leq m \}$ to get $\hat{t}$ which is equivalent to using $\{\hat{\psi}(t_i) | 1 \leq i \leq m\}$. This yields
$$
|\hat{t}-t^*| \leq  \left( \frac{4 \sup_{t \in [0,T]} |- \Tilde{\hat{{\psi}}}(t) - \psi(t)|}{|\Delta|} + 2\rho \right) \frac{T}{ min\{T-t^*,t^*\} }.
$$ Combining the previous two results, we conclude that $$
|\hat{t}-t^*| \leq  \left( \frac{4 \min_{w \in \{\pm1\}  } \sup_{t \in [0,T]} | \Tilde{\hat{{\psi}}}(t) - w\psi(t)|}{|\Delta|} + 2\rho \right) \frac{T}{ min\{T-t^*,t^*\} }.
$$

\end{proof}

\section{Verification of KKT condition}\label{Sec:append-KKT}

First we recall the Karush-Kuhn-Tucker (KKT) conditions for the minimizer of the constrained optimization problem, stated below:

\begin{theorem}[Necessary conditions for the minimizer]\label{thm:KKT}
Consider the constrained optimization problem: 
\begin{align*}
\min ~  & f(\mathbf{x}) \\
\textrm{subject to }   g_1(\mathbf{x}) &\leq 0, \cdots,
g_k(\mathbf{x}) \leq 0
\end{align*}
Suppose the minimizer of this problem is $\mathbf{x}^*$. If $g_{i_1}(\mathbf{x}^*)=g_{i_2}(\mathbf{x}^*)=\cdots=g_{i_j}(\mathbf{x}^*)=0$ (namely, the constraints are active), and $g_l(\mathbf{x}^*)<0$ for $l \notin \{i_1,\cdots,i_j\}$, then if the gradients of the active constraints evaluated at $\mathbf{x}^*$ are linearly independent, there exist $\lambda_1,\cdots,\lambda_j \geq 0$ such that:    
\begin{equation}\label{equ:KKT}
-\nabla f(\mathbf{x}^*)=\sum^{j}_{s=1} \lambda_s \nabla g_{i_s}(\mathbf{x}^*).
\end{equation}
\end{theorem}

Our optimization problem of interest is given by:
$$\min_{\alpha',\beta_L',\beta_R', \beta_L' \neq \beta_R'} \max \{ |\alpha'|,|\alpha'-\beta_L'\hat{t}|, |\alpha'+\beta_R'(t^*-\hat{t})|, |\alpha'+\beta_R'(T-\hat{t})-|\Delta|(T-t^*) | \}.$$
Therefore, without loss of generality, we may take $\Delta>0$, and our optimization problem can be written as: 
$$
\min_{\alpha,\beta_L,\beta_R, \beta_L \neq \beta_R} \max \{ |\alpha|,|\alpha-\beta_L\hat{t}|, |\alpha+\beta_R(t^*-\hat{t})|, |\alpha+\beta_R(T-\hat{t})-\Delta(T-t^*) | \}.
$$

This is equivalent to:
\begin{align*}
\min  ~ f(z,\alpha,\beta_L,\beta_R)&=z; \\
\text{subject to } 
g_1&=\alpha-z  \leq  0, \\
g_2&=-\alpha-z  \leq  0, \\
g_3&=\alpha-\beta_L\hat{t}-z  \leq 0, \\
g_4&=-\alpha+\beta_L\hat{t}-z  \leq 0 ,\\
g_5&=-\alpha-\beta_R(t^*-\hat{t})-z \leq 0, \\
g_6&=\alpha+\beta_R(t^*-\hat{t})-z  \leq 0, \\
g_7&=-\alpha-\beta_R(T-\hat{t})+\Delta(T-t^*)-z  \leq 0, \\
g_8&=\alpha+\beta_R(T-\hat{t})-\Delta(T-t^*)-z  \leq 0.
\end{align*}

We recall $\Delta > 0, 0<\hat{t}<t^*<T.$ Then 
$$
\nabla f(z,\alpha,\beta_L,\beta_R)= (1,0,0,0)^{\top}.
$$

Because $\nabla f(z,\alpha,\beta_L,\beta_R)$ never vanishes, the minimizer $\mathbf{x}^*$ must be on the boundary of the feasible set, so at least one of the constraints is active: namely, there exists $i$ with $g_i(\mathbf{x}^*)=0$. In the following analysis, we enumerate all possible sets of active constraints and find the only feasible ones, which yields the minimizer. 

For ease of notation, we form a matrix $\nabla$ with its $i$th column to be\\$\nabla g_i(z,\alpha,\beta_L,\beta_R)$:
$$
\nabla=\left[\begin{array}{cccccccc}-1 & -1 & -1 & -1 & -1 & -1 & -1 & -1 \\ 
1 & -1 & -1 & 1 & -1 & 1 & -1 & 1\\ 
0 & 0 & \hat{t} & -\hat{t} & 0 & 0 & 0 & 0 
\\ 0 & 0 & 0 & 0 & -\left(t^*-\hat{t}\right) & t^*-\hat{t} & -(T-\hat{t}) & T-\hat{t}\end{array}\right].
$$

Also if only one of the constraints are active at the minimizer, then by KKT conditions Theorem \ref{thm:KKT}, there exists a $\lambda \geq 0$ such that:
$$
\lambda \nabla g_i=(-1,0,0,0)^{\top}.
$$

This equation has no solution for any $i \in \{1,2,\cdots,8\}$. Thus at least two of constraints are active at the minimizer. In addition, we note if any one of the four pairs of $(g_1,g_2),(g_3,g_4)$,$(g_5,g_6),(g_7,g_8)$ are active, then $z=0$, then $\Delta=0$, which violates our assumption. Thus in none of these four pairs can both constraints be active. 

For $g_3$ and $g_4$, because $\hat{t} > 0$ and only $g_3$ and $g_4$ have the third entry non-zero, to satisfy the equation \ref{equ:KKT}, the coefficients of $\nabla g_3$ and $\nabla g_4$ have to be equal. Since at most one of these coefficients is nonzero, they must both be zero.

Now we consider the case in which the active constraints are among $g_5,$ $g_6,$ $g_7,$ $g_8$. Suppose the only constraints that are active are $(g_5,g_7)$. (Note that it is possible that $g_3$ or $g_4$ is also active, but we have shown the corresponding coefficient is zero, and hence this cannot influence the equation \ref{equ:KKT}.) In this case, the equation \ref{equ:KKT} has no solution. Similarly,  $(g_6,g_8)$ cannot be the only active constraints. Note that this still holds if one of $g_1, g_2$ is active. Then for $(g_5,g_8)$ to satisfy the equation \ref{equ:KKT}, we find $T=t^*$ which again violates our assumption. A similar analysis holds for $(g_6,g_7)$.
The above analysis shows that one of $g_1$ and $g_2$ has to be active, that is $z=\alpha$ or $z=-\alpha$. 

Now if $(g_2,g_5,g_8)$ are all active, then we have:
$$
z=-\alpha=-\alpha-\beta_R(t^*-\hat{t})=\alpha+\beta_R(T-\hat{t})-\Delta(T-t^*)
$$

Thus $\beta_R=0$ and $z=\frac{\Delta(t^*-T)}{2}<0$, and this is infeasible. 

For $(g_1,g_5,g_8)$ we have:
$$
z=\alpha=-\alpha-\beta_R(t^*-\hat{t})=\alpha+\beta_R(T-\hat{t})-\Delta(T-t^*)
$$

Thus $\alpha=-\frac{\Delta(t^*-\hat{t})(T-t^*)}{2(T-\hat{t})}$, $z=-\frac{\Delta(t^*-\hat{t})(T-t^*)}{2(T-\hat{t})}<0$, which is infeasible.

For $(g_1,g_6,g_7)$, by solving the active constraints similarly we have $z=\alpha=\frac{\Delta(T-t^*)}{2}$, which is feasible. Then we check the KKT condition \ref{equ:KKT}:

\begin{align*}
\begin{cases}
    - \lambda_1-\lambda_6-\lambda_7=-1 \\
    \lambda_1+\lambda_6-\lambda_7=0 \\
    \left(t^*-\hat{t}\right)\lambda_6-\left(T-\hat{t}\right) \lambda_7=0
\end{cases}
\end{align*}

Solving this, we note $\lambda_1=\frac{1}{2}\left(1- \frac{T-\hat{t}}{t^*-\hat{t}} \right) < 0$, which violates the KKT conditions. 

For $(g_2,g_6,g_7)$ we have:
$$
z=-\alpha=\alpha+\beta_R(t^*-\hat{t})=-\alpha-\beta_R(T-\hat{t})+\Delta(T-t^*)
$$

Thus $\alpha=-\frac{\Delta(t^*-\hat{t})(T-t^*)}{2(T-\hat{t})}$, $z=\frac{\Delta(t^*-\hat{t})(T-t^*)}{2(T-\hat{t})}$. We next check KKT conditions by solving the equations:

\begin{align*}
\begin{cases}
    -\lambda_2-\lambda_6-\lambda_7=-1 \\
    -\lambda_2+\lambda_6-\lambda_7=0 \\
    \left(t^*-\hat{t}\right)\lambda_6-\left(T-\hat{t}\right) \lambda_7=0
\end{cases}
\end{align*}

We obtain the following solutions:
\begin{align*}
\begin{cases}
    \lambda_2= \frac{1}{2}\left(1- \frac{t^*-\hat{t}}{T-\hat{t}}\right)>0 \\
    \lambda_6=\frac{1}{2} > 0 \\
    \lambda_7=\frac{t^*-\hat{t}}{2\left(T-\hat{t}\right)} >0 
\end{cases}
\end{align*}

This satisfies the KKT condition, and the corresponding gradients are linearly independent: 
$$
\left[\begin{array}{ccc}-1 & -1 & -1  \\ 
-1 & 1 & -1 \\ 
0 & 0 & 0 \\ 
0 & t^*-\hat{t} & -(T-\hat{t}) 
\end{array}\right]
$$

Because these active constraints at the minimizer are the only set that satisfies the KKT conditions (with the possible addition of $g_3$ or $g_4$), these active constraints yield that $z=\frac{\Delta\left(t^*-\hat{t}\right)(T-t^*)}{2\left(T-\hat{t}\right)}$. We conclude the minimum value of this optimization problem is $\frac{\Delta\left(t^*-\hat{t}\right)(T-t^*)}{2\left(T-\hat{t}\right)}$. 

\section{Proof of Theorem~\ref{thm:consistency asy1-d}}
\begin{proof}
We apply Lemma~\ref{lem:consistency}, then bound the error $\sup_{t\in[0,T]} \left|\tilde{\hat{\psi}}^{(m)}_1(t)-w_m\psi_0(t)\right|$. By Lemma~\ref{Lem:endpointlemma}, this is at most $\max\{|\hat{\psi}_1^{(m)}(t)-w_m\psi_0(t)|:t\in\mathcal{T}\}$. We bound this as
$$\max_{t\in\mathcal{T}} |\hat{\psi}_1^{(m)}(t)-w_m \psi_0(t)|\leq \max_{t\in\mathcal{T}}|\hat{\psi}_1^{(m)}(t)-w \psi_1^{(m)}(t)|+\max_{t\in\mathcal{T}}|w\psi_1^{(m)}(t)-w_m\psi_0(t)|.$$
By Corollary~\ref{Cor:3}, under the assumption of asymptotic Euclidean 1-realizability, we have a bound on the first quantity given by 
$$\max_{t\in\mathcal{T}}|\hat{\psi}_1^{(m)}(t)-w \psi_1^{(m)}(t)|\leq C\left(\frac{m\log(n)}{\sqrt{n}\sqrt{\lambda_1(E_{\varphi})}}+\sqrt{\frac{\sum_{i=2}^m \lambda_i^2(E_{\varphi})}{\lambda_1(E_{\varphi})}}\right).$$ We can apply Lemma~\ref{Lem:endpointlemma} once again to show $$\max_{t\in\mathcal{T}} |\psi_1^{(m)}(t)-w_m\psi_0(t)| = \sup_{t\in[0,T]}|\tilde{\psi}_1^{(m)}(t)-w_m\psi_0(t)|,$$ which proves the theorem.
\end{proof}

\section{Proof of Corollary \ref{Cor:uniform time step result}}
We wish to apply Lemma~\ref{lem:consistency}, and obtain sharper results for the error $$\sup_{t\in[0,T]}|\tilde{\hat{\psi}}(t)-w_m\psi(t)|$$ in the exact Euclidean realizable setting, where $\psi(t)$ is the exact mirror. We will do this in the $\alpha$-H\"{o}lder case first (Theorem~\ref{thm:sup controll}), then specialize to the case where the mirror takes the form $\psi(t)=at+b+\Delta(t-t^*)I_{\{t>t^*\}}$ in Theorem~\ref{Thm:exact 1-d consistency}. We first prove a lemma concerning the approximation of an $\alpha$-H\"{o}lder function from the linear interpolation of its sampled points:
\begin{lemma}\label{lem:tilde-0}
    Assume that $\psi(t)$ is an $\alpha$-H\"{o}lder continuous with constant $L$ function on $t \in [0,T]$. Let the sampled times $\{t_1,t_2,...t_{m}\} \subset [0,T]$ be given and assume $t_1=0$ and $t_{m}=T$. Denote $$\mathop{max}\limits_{1\leq i < m} |t_i-t_{i+1}|^\alpha=\rho.$$ If $\Tilde{\psi}(t)$ is the piecewise linear interpolation between the points\\ $\left\{\left(t_i, \psi\left(t_i\right)\right), 1 \leq i \leq m \right\}$, then
    $$
    \sup_{t \in [0,T]} |\Tilde{\psi}(t)-\psi(t)| \leq \rho L.
    $$
\end{lemma}
\begin{proof}
For $t \in [0,T]$ and $t \neq t_k$ for $1 \leq k \leq m $, because we assume $t_1=0$ and $t_m=T$, we can always find $t_i$ and $t_{i+1}$ where $1 \leq i < m $ such that  $t \in (t_i,t_{i+1})$. Then there exists a $0 < \lambda < 1$ such that $t=\lambda t_i+ (1-\lambda) t_{i+1}$. Since $\Tilde{\psi}(t)$ is the piecewise linear interpolation between the points $\{(t_i, \psi(t_i)), 1 \leq i \leq m \}$, it follows that $$\Tilde{\psi}(t)= \lambda \psi(t_i) +(1-\lambda) \psi(t_{i+1}).$$
We thus have
\begin{align*}
|\Tilde{\psi}(t)-\psi(t)| &= | \lambda (\psi(t_i)-\psi(t)) +(1-\lambda) (\psi(t_{i+1})- \psi(t)) |  \\
& \leq \lambda L|t-t_i|^{\alpha} + (1-\lambda)L|t_{i+1}-t|^{\alpha}\\
& \leq (\lambda L+ (1-\lambda)L)|t_{i+1}-t_i|^{\alpha} \\
& = L|t_i-t_{i+1}|^{\alpha}.
\end{align*}
Thus 
$$
\sup_{t \in [0,T]}|\Tilde{\psi}(t)-\psi(t)|   \leq  L \max\limits_{1\leq i < m} |t_i-t_{i+1}|^\alpha = \rho L.
$$
\end{proof}
Now we provide our bound on the sup-norm error for the estimation of the mirror in the $\alpha$-H\"{o}lder case.
\begin{theorem} \label{thm:sup controll}
    Consider an exactly Euclidean $1$-realizable latent position process with $\alpha$-H\"{o}lder mirror $\psi(t)$, where $L$ denotes the H\"{o}lder constant on $t \in [0,T]$. Let $\{\hat{\psi}(t_i) | 1 \leq i \leq m \}$ denote the estimated mirror generated from the time series of graphs on $n$ vertices with this LPP. Denote by $\Tilde{\hat{\psi}}(t)$ the piecewise linear interpolation between the points $\left\{\left(t_i, \hat{\psi}(t_i)\right) | 1 \leq i \leq m\right\}$.
    
    Assume $t_1=0$, $t_m=T$, $\max_{1\leq i \leq m} |t_i-t_{i+1}|^\alpha=\rho$ and \break \hbox{$\lambda_1\left(-\frac{1}{2}P\mathcal{D}^{(2)}P\right) \sim \Theta(m)$} where $\mathcal{D}^{(2)}$ is the $m \times m$ entrywise square of the distance matrix of the mirror $\psi(t)$, namely $\mathcal{D}^{(2)}_{ij}=\|\psi(t_i)-\psi(t_j)\|^2$ 
    and $P=I-J/m$, where $J$ is the $m \times m$ matrix with all entries equal to 1.  
    Then there is a constant $C$ and a $w_m\in\{\pm1\}$ such that with high probability,
    $$
    \sup_{ t \in [0,T] } |\Tilde{\hat{\psi}}(t)-w_m\psi(t)| \leq  \rho L + C \frac{ m \log(n)}{\sqrt{n} \sqrt{\lambda_1\left(-\frac{1}{2}P\mathcal{D}^{(2)}P\right)} }.
    $$
\end{theorem}
\begin{proof}
By Lemmas~\ref{Lem:endpointlemma} and \ref{lem:tilde-0}, we know there exists a $w_m\in\{\pm1\}$ such that
\begin{align*}
\sup_{ t \in [0,T] } |\Tilde{\hat{\psi}}(t)-w_m\psi(t)| &\leq \sup \limits_{t \in [0,T]} |\Tilde{\hat{\psi}}(t)-w_m\Tilde{\psi}(t)| + \sup_{t \in [0,T]} |w_m\Tilde{\psi}(t)-w_m\psi(t)|, \\
& \leq  \max_{1\leq i \leq m} | \hat{\psi}(t_i)-w_m\psi(t_i) | + \rho L,\\
& \leq \frac{C m \log(n)}{\sqrt{n} \sqrt{\lambda_1(-\frac{1}{2}P\mathcal{D}^{(2)}P)}}+\rho L,
\end{align*}
where the last bound follows by Corollary \ref{cor:simple_upperbound} and exact
1-realizability. This completes the proof.
\end{proof}

Now we specialize this result to the case where the mirror is piecewise linear.
\begin{theorem}\label{Thm:exact 1-d consistency}
Suppose the setting of Theorem~\ref{thm:sup controll}, but further suppose that the mirror is given by $\psi(t) = at+b+\Delta(t-t^*)I_{\{t>t^*\} } ~ t \in [0,T]$, $\Delta \neq 0$, and $t^*\in (t_1,t_m)$. Then there exists a constant $C$ such that with high probability: 

$$
|\hat{t}-t^*| \leq  \left( \frac{C m\log(n)}{ \sqrt{\Sigma^{m}_{i=1}\psi_{centered}^2 \left(t_i \right)} |\Delta|\sqrt{n} }   + 6\rho \right) \frac{T}{ min\{T-t^*,t^*\} } .
$$      
\end{theorem}
\begin{proof}
The true mirror $\psi$ has the form $\psi(t) = at+b+\Delta(t-t^*)I_{\{t>t^*\}}$ for $t \in [0,T]$, so $\psi(t)$ is Lipschitz continuous with constant $|\Delta|$. We can now apply Lemma~\ref{lem:consistency}, then bound the error in approximating the mirror with Theorem~\ref{thm:sup controll}. We need to compute $\lambda_1\left(-\frac{1}{2}P\mathcal{D}^{(2)}P\right)$. Let $\psi(t) = at+b+\Delta(t-t^*)I_{\{t>t^*\} }$ for $ t \in [0,T]$. Define $\mathcal{D}^{(2)}$ to be the matrix whose $i,j$th entry is
$\mathcal{D}^{(2)}_{ij}=|\psi(t_i)-\psi(t_j)|^2$ for $t_i$ and  $t_j$ belonging to $ \{t_1, \cdots, t_m\}$.
Put $c=\frac{1}{m}\sum_{i=1}^m\psi(t_i)$ and
$\psi_{centered}(t)=\psi(t)-c$.
Denote by $\Psi_m$ the $m \times 1$ vector
$$\Psi_m=\left(\psi_{centered}(t_1),\psi_{centered}(t_2),\cdots,\psi_{centered}(t_m)\right).$$ 
Note that
$-\frac{1}{2}P\mathcal{D}^{(2)}P= \Psi_m \Psi^{\top}_m
$, and hence 
$$\lambda_1\left(-\frac{1}{2}P\mathcal{D}^{(2)}P\right)=\lambda_1\left(\Psi_m \Psi^{\top}_m\right)=\Psi^{\top}_m \Psi_m=\Sigma^{m}_{i=1}\psi_{centered}^2 \left(t_i \right)$$
as required.
\end{proof}
We now have all of the ingredients required to prove Corollary~\ref{Cor:uniform time step result}.
\begin{proof}
We start from the bound in Theorem~\ref{Thm:exact 1-d consistency}. In this case $\rho=\frac{T}{m-1}$ and $t_i=\frac{\left(i-1\right)T}{m-1}$ for $1 \leq i \leq m$. We require a lower bound on the denominator of the first term. Set $c_m=\frac{1}{m}\sum_{i=1}^m \psi(t_i)$. By the Riemann integrability of $\psi(t)$,  $c_m \to \frac{1}{T}\int_0^T \psi(t)\mathrm{d}t$ as $m \to \infty$. 
Thus, we observe that
\begin{align*}
\frac{1}{m}\sum^m_{i=1}\psi_{centered}^2(t_i)&= \frac{1}{m}\sum_{i=1}^m \left(\psi(t_i)-c_m\right)^2\\
&=\frac{1}{m}\sum_{i=1}^m \left( \psi(t_i)^2-2\psi(t_i)c_m+c_m^2 \right)\\
&=\left\{\frac{m-1}{mT}\sum^m_{i=1}\frac{T}{m-1}\psi^2\left(\frac{\left(i-1\right)T}{m-1}\right)\right\} -c_m^2\\
&\to \frac{1}{T} \int_0^T\psi^2(t) \mathrm{d}t - \left(\frac{1}{T}\int_0^T \psi(t)\mathrm{d}t\right)^2.
\end{align*}

Since $\psi(t)=at+b+\Delta(t-t^*)I_{\{t>t^*\}}$, these integrals can be computed explicitly. The value of the integral is clearly independent of $b$, so without loss of generality we set $b=0$. Then we obtain
\begin{align*}
\int_0^T \psi(t)\,\mathrm{d}t&= a\frac{T^2}{2}+\Delta\frac{(T-t^*)^2}{2},\\
\int_0^T \psi^2(t)\,\mathrm{d}t&= a^2\frac{T^3}{3}+a\Delta(T-t^*)^2\frac{2T+t^*}{3}+\Delta^2\frac{(T-t^*)^3}{3}.
\end{align*}
So the limit equals 
\begin{align*} 
\frac{1}{T}\int_0^T\psi^2(t)\,\mathrm{d}t&-\left(\frac{1}{T}\int_0^T\psi(t)\,\mathrm{d}t\right)^2\\
&=a^2\frac{T^2}{12}+a\Delta\frac{(T-t^*)^2}{2T}\left(\frac{T+2t^*}{3}\right)+\Delta^2\frac{(T-t^*)^3}{3T^2}\left(\frac{T+3t^*}{4}\right)\\
&\geq \frac{1}{12T}\left(a^2T^3+\Delta^2(T-t^*)^3\right).
\end{align*}

\end{proof}

\section{Beyond the first dimension}
\label{Sec:beyond-1st-dim}
 
The consistency results discussed previously hinge on the assumption that the mirror is asymptotically piecewise linear in the first dimension, with the remaining dimensions carrying negligible signal. However, as observed in Figure  \ref{fig:cmds_changepoint}, the changepoints are also detectable in higher dimensions. To see this more precisely, we revisit the distance matrix for Model \ref{changepointmodel}. Recall in Lemma \ref{lem:dmv-changepoint},
$$
\left(\mathcal{D}^{(2)}_{\varphi_m}\right)_{i,j}=
\begin{cases}
\frac{p^2}{m^2}(i-j)^2+\frac{p-p^2}{m^2}|i-j| &\quad i,j<t_m^* \\
\left(\frac{p}{m}(t_m^*-i)+\frac{q}{m}(j-t_m^*) \right)^2 &\\
\quad +\frac{p-p^2}{m^2}(t_m^*-i)+\frac{q-q^2}{m^2}(j-t_m^*) &\quad i<t_m^*<j \\
\frac{q^2}{m^2}(j-i)^2+\frac{q-q^2}{m^2}|i-j| &\quad t_m^*<i,j.
\end{cases}
$$ To study the mirror of Model \ref{changepointmodel}, we perform CMDS on this distance matrix, which amounts to taking an eigendecomposition of $-\frac{1}{2}P\mathcal{D}^{(2)}_{\varphi_m}P$. When $m$ is large, this eigendecomposition can be approximated through the kernel method of \cite{diaconis2008horseshoes}, in which we consider a kernel function $\kappa^m(x,y)$ that provides a continuous-time analogue of the discrete distance matrix, and a double-centering operator that approximates the discrete centering with an integral. 

To that end, set $t^*:=\frac{t_m^*}{m}$. Define the corresponding kernel function  $\kappa^m(x,y):[0,1]^2 \to \R$ as follows:
\begin{equation*}
\kappa^m(x,y)=
\begin{cases}
p^2(x-y)^2+\frac{p-p^2}{m}|x-y| & \quad x,y<t^*. \\
\\
\left(t^*-x\right)^2p^2+\frac{p-p^2}{m}\left(t^*-x\right)\\
+ \left(y-t^*\right)^2q^2 +\frac{q-q^2}{m}\left(y-t^*\right)\\
+2pq\left(y-t^*\right)\left(t^*-x\right) & \quad x<t^*<y. \\
\\
\kappa^m(y,x) & \quad y<t^*<x .\\
\\
q^2(x-y)^2+\frac{q-q^2}{m}|x-y| & \quad t^*<x,y. \\
\end{cases}    
\end{equation*} and a double-centering operator $\mathcal{DC}$:
$$
\mathcal{DC}[\kappa(x,y)]:= \kappa(x,y)-\int_0^1\kappa(x,y)\,\mathrm{d}x -\int_0^1\kappa(x,y)\,\mathrm{d}y + \int_0^1 \int_0^1\kappa(x,y) \,\mathrm{d}x \mathrm{d}y.
$$ 
The continuous-time analogue of $P\mathcal{D}^{(2)}_{\varphi_m}P$ is 
$\mathcal{DC}[\kappa^m(x,y)]$, defined by

\begin{equation}\label{equ:kappa_ct_2}
\mathcal{DC}[\kappa^m(x,y)]=
\begin{cases}
p^2\left( \left(2t_m^*-{t_m^*}^2\right)(x+y)-2xy\right) \\
+\frac{p-p^2}{m}\left(|x-y|-x^2-y^2+x+y\right)\\
+ pq\left(1-{t_m^*}^2\right)(x+y)  +c_1, & x,y<t_m^* ;
\\
\\
-p^2{t_m^*}^2x-\frac{p-p^2}{m}x^2\\
+ q^2({t_m^*}-1)^2y 
+ \frac{q-q^2}{m}(2y-y^2) \\
+ pq\left(-2xy+\left(2{t_m^*}-{t_m^*}^2\right)y\right)\\
+pq\left(-{t_m^*}^2+2{t_m^*}+1\right)x+c_2,  & x<t_m^*<y; 
\\
\\
\mathcal{DC}[\kappa^m(y,x)] & y<{t_m^*}<x; 
\\
\\
q^2\left( \left(2t_m^*-{t_m^*}^2\right)(x+y)-2xy\right) \\
+\frac{q-q^2}{m}\left(|x-y|-x^2-y^2+x+y\right)\\
+ pq(1-{t_m^*}^2)(x+y)  +c_3, & x,y>t_m^*;
\end{cases}
\end{equation}
where $c_1, c_2, c_3$ are constants.

Observe that each of these kernels induces an associated integral operator $I_{\kappa}:L^2\left( [0,1] \right) \rightarrow L^2\left( [0,1] \right) $ defined by
$$I_{\kappa}[f](x)=\int_{0}^1 \kappa(x,y) \, f(y) dy.$$

The $\kappa^m_{c}$ kernel defined as $\kappa^m_c(x,y):= -\frac{1}{2}\mathcal{DC}[\kappa^m(x,y)]$  approximates our doubly-centered matrix in that the entrywise errors decay with $m$, as the following lemma demonstrates.

\begin{lemma}\label{lem:kappa_ct_2}
With $\kappa^m_{c}$ defined as $\kappa^m_c(x,y):= -\frac{1}{2}\mathcal{DC}[\kappa^m(x,y)]$, 
define a matrix $S$ with its $i,j$ entry as:
$$
S_{i,j}:= \kappa^m_{c}\left(\frac{i}{m},\frac{j}{m}\right).
$$
Then for large $m$ we have
$$
\max_{1\leq i \leq m,1 \leq j \leq m} 
\bigg| S_{i,j} - \left(-\frac{1}{2}P\mathcal{D}^{(2)}_{\varphi_m}P\right)_{i,j} \bigg| \sim \Theta\left(\frac{1}{m}\right).
$$
\end{lemma}

This lemma is proved in Appendix \ref{sec:lemproof}. For an integral operator $I_{\kappa}$, its eigenvalues $\lambda$ and eigenfunctions $f \in L^2([0,1])$ satisfy
$$
I_{\kappa}[f](x)=\lambda f(x)\text{~ for all } x \in [0,1].
$$

The following theorem gives an explicit form for the eigenfunctions of the integral operator $I_{\kappa^m_c}$.

\begin{theorem}\label{thm:changepoint eigenfunction}
Let the kernel function $\kappa^m_{c}(x,y)=-\frac{1}{2}\mathcal{DC}[\kappa^m(x,y)]$, where $\mathcal{DC}[\kappa^m(x,y)]$ is defined in Eq (\ref{equ:kappa_ct_2}). Suppose that the eigenvalues of its corresponding integral $I_{\kappa_c^m}$ operator are positive. Then any eigenfunction of $I_{\kappa^m_c}$ must take the following form:
\begin{align}
k(x)&=\begin{cases}
 A_\pre\cos\left(\sqrt{\frac{p(1-p)}{m\lambda}}x\right)+B_\pre\sin\left(\sqrt{\frac{p(1-p)}{m\lambda}}x\right) \text{\quad for $x \leq t^{*}$}\\
 A_\post\cos\left(\sqrt{\frac{q(1-q)}{m\lambda}}x\right)+B_\post\sin\left(\sqrt{\frac{q(1-q)}{m\lambda}}x\right) \text{\quad for $x > t^{*}$},
\notag
\end{cases}
\end{align}
where $\lambda$, $A_\pre$, $B_\pre$ and $A_\post$, $B_\post$ are real numbers and satisfy $\int_0^1 k(t) \mathrm{d}t=0$, $\int_0^1 k^2(t) \mathrm{d}t =1 $.
Thus, the approximated mirror for \ref{changepointmodel} will be $k(x)$ scaled by the square root of the eigenvalues. 
\end{theorem}

Determining $\lambda$, $A_\pre$, $B_\pre$ and $A_\post$, $B_\post$ requires the solution of an especially complicated equation, but an explicit solution is unnecessary. Rather, Theorem \ref{thm:changepoint eigenfunction} demonstrates that there is a change in frequency at $t^*$ in all eigenfunctions, making the estimated eigenvectors in all dimensions a useful object of study for changepoint localization.

As shown in Figure \ref{fig:cmds_changepoint_with_blut_dots}, the first dimension of the estimated mirror exhibits approximate piecewise linearity with a slope change at $t^*$. The second and third dimensions also exhibit a change in behavior at $t^*$, where the frequency of the cosine and sine curves change abruptly. This suggests a robustness to misspecification of embedding dimension: if we choose too large a dimension for the estimated mirror, the corresponding eigenfunctions still buttress the case for a changepoint at $t^*$, but estimates of these eigenfunctions become increasingly variable. This bias-variance tradeoff is illustrated in Section~\ref{sec:numerical}. 

\subsection{Proof of Lemma \ref{lem:kappa_ct_2}}
\label{sec:lemproof}
\begin{proof}
Consider the function $\kappa^m(x,y)$ as  defined in Appendix \ref{Sec:beyond-1st-dim}. Note this function is bounded on $[0,1] \times [0,1]$. Putting $x_i=i/m$, we observe $$\kappa^m(x_i,x_j)=(\mathcal{D}^{(2)}_{\varphi_m})_{i,j}.$$ Since $P=I-\frac{J}{m}$ and $J$ is an all ones matrix,
$P\mathcal{D}^{(2)}_{\varphi_m}P=\mathcal{D}^{(2)}_{\varphi_m}-\frac{1}{m}J\mathcal{D}^{(2)}_{\varphi_m}-\frac{1}{m}\mathcal{D}^{(2)}_{\varphi_m}J+\frac{1}{m^2}J\mathcal{D}^{(2)}_{\varphi_m}J$. 

Recall that when approximating the Riemann integral by the Riemann sum computed at the right endpoint of a given partition, the following bound holds for the error. 
\begin{lemma}[Right hand rule error bound for numerical integration]
\label{Lem:right hand rule}
Suppose that $f(x)$ is differentiable on $[a,b]$ and $|f^{'}(x)| \leq K_1$. Then
$$
\int^b_a f(x) \,\mathrm{d}x= \sum^m_{i=1}\frac{b-a}{m} f\left(\frac{i}{m}\right) +E_R
\quad 
\text{where} \quad |E_R|\leq \frac{K_1(b-a)^2}{2m}. 
$$
\end{lemma}
We find
$$
\left(\frac{1}{m}J\mathcal{D}^{(2)}_{\varphi}\right)_{i,j}
=\sum^m_{t=1}\frac{1}{m}\left(\mathcal{D}^{(2)}_{\varphi}\right)_{t,j}
=\sum^m_{t=1}\frac{1}{m}\kappa^m\left(\frac{t}{m},x_j\right),
$$
and
\begin{align*}
\sum^m_{t=1}\frac{1}{m}\kappa^m\left(\frac{t}{m},x_j\right) &= \sum^{t^*_m}_{t=1}\frac{1}{m}\kappa^m\left(\frac{t}{m},x_j\right) + \sum^m_{t=t^*_m+1}\frac{1}{m}\kappa^m\left(\frac{t}{m},x_j\right) \\ 
&= \int_0^{t^*} \kappa^m(x,x_j)dx  + R_{x_j}+   \int_{t^*}^{1} \kappa^m(x,x_j)dx  + R'_{x_j},
\end{align*}
where $R_{x_j}$ is the error in approximating the first integral by the first sum, and $R'_{x_j}$ is the error in approximating the second integral by the second sum.

Note here we apply the right hand rule error bound because $\kappa^m(x,y)$ is differentiable with respect to $x$ on $[0,t^*]$ for any fixed $y \in [0,1]$ and differentiable with respect to $x$ on $[t^*,1]$ for any fixed $y \in [0,1]$. In addition, the derivative $\frac{d\kappa^m(x,y)}{dx}$ on $x \in [0,t^*]$ is bounded for any $y \in [0,1]$; the same holds for $\frac{d\kappa^m(x,y)}{dx}$ on $x \in [t^*,1]$. Thus by Lemma \ref{Lem:right hand rule} we conclude that 
\begin{align*}
R_{x_j} &\leq \frac{\max_{x \in [0,t^*]}|\frac{d\kappa^m(x,y)}{dx}|}{2m}\\
&\leq \frac{ \max_{y \in [0,1]} \max_{x \in [0,t^*]}|\frac{d\kappa^m(x,y)}{dx}|}{2m}\\
&\leq \frac{C}{m} \text{~ for ~}  x_j \in [0,1].
\end{align*}
The analogous bound holds for $R'_{x_j}$. Together, we have an uniform upper bound for the error in the right-hand Riemann sum; namely for every $x_j \in [0,1]$,
$$
\sum^m_{t=1}\frac{1}{m}\kappa\left(\frac{t}{m},x_j\right)= \int_0^1 \kappa(x,x_j)dx  + R^{(1)}  \quad 
\text{where} \quad  \left|R^{(1)}\right|  \sim O\left(\frac{1}{m}\right).
$$
Note that
$$
\left(\frac{1}{m}\mathcal{D}^{(2)}_{\varphi}J\right)_{i,j}
=\sum^m_{t=1}\frac{1}{m}\left(\mathcal{D}^{(2)}_{\varphi}\right)_{i,t}
=\sum^m_{t=1}\frac{1}{m}\kappa\left(x_i,\frac{t}{m}\right),
$$
and since $\kappa^m(x,y)$ is symmetric, we have that for every $x_i \in [0,1]$,
$$
\sum^m_{t=1}\frac{1}{m}\kappa^m\left(x_i,\frac{t}{m} \right) = \int_0^1 \kappa^m(x_i,x)dx  + R^{(2)}
\quad 
\text{where} \quad  \left|R^{(2)}\right| \sim O\left(\frac{1}{m}\right). 
$$
Now recall that
$$
\left(\frac{1}{m^2}J\mathcal{D}^{(2)}_{\varphi_m}J\right)_{i,j}
= \sum^m_{t'=1}\sum^m_{t=1}\frac{1}{m^2}\left(\mathcal{D}^{(2)}_{\varphi_m}\right)_{t',t}
= \sum^m_{t'=1}\sum^m_{t=1}\frac{1}{m^2}\kappa\left(\frac{t'}{m},\frac{t}{m}\right),
$$
and thus 
\begin{align*}
\sum^m_{t'=1}\sum^m_{t=1}\frac{1}{m^2}\kappa^m\left(\frac{t'}{m},\frac{t}{m}\right) 
&=\sum^m_{t'=1} \frac{1}{m}\left( \sum^m_{t=1} \frac{1}{m} \kappa^m\left(x_{t'},\frac{t}{m}\right)\right) \\
&= \sum^m_{t'=1} \frac{1}{m}\left(   \int_0^1\kappa(x_{t'},y)dy + R^{(2)}   \right) \\
&=\sum^m_{t'=1} \frac{1}{m}   \int_0^1\kappa^m(x_{t'},y)dy + \sum^m_{t'=1} \frac{1}{m} R^{(2)} \\
& =\int_0^1\int_0^1\kappa(x,y)dxdy + R^{(3)} +  R^{(2)}\\
\text{where} \quad \left|R^{(3)}\right| &\leq \frac{1}{2m}\max_ {x\in[0,1]} \left|\frac{d}{dx}\int_0^1\kappa^m(x,y)dy\right| \sim O\left(\frac{1}{m}\right).
\end{align*}
Then for any $i,j \in \{1,2,\cdots,m\}$:
\begin{align*}
\left(-\frac{1}{2}P\mathcal{D}^{(2)}_{\varphi_m}P\right)_{i,j} 
&=-\frac{1}{2}\left(\mathcal{D}^{(2)}_{\varphi_m}-\frac{1}{m}J\mathcal{D}^{(2)}_{\varphi_m}-\frac{1}{m}\mathcal{D}^{(2)}_{\varphi_m}J+\frac{1}{m^2}J\mathcal{D}^{(2)}_{\varphi_m}J\right)_{i,j} \\
&=-\frac{1}{2}\bigg[
\kappa^m(x_i,x_j)-\int_0^1\kappa^m(x,x_j)dx \\
& \qquad-\int_0^1\kappa^m(x_i,y)dy+\int_0^1\int_0^1\kappa^m(x,y)dxdy \\
&\qquad-R^{(1)}
-R^{(2)}
+R^{(3)}
+R^{(2)}
\bigg]. 
\end{align*}
Because 
$$
\adjustbox{width=4.513in}{$\displaystyle \kappa^m_{c}(x_i,x_j)=-\frac{1}{2}\left[
\kappa(x_i,x_j)-\int_0^1\kappa^m(x,x_j)dx-\int_0^1\kappa(x_i,y)dy+\int_0^1\int_0^1\kappa(x,y)^mdxdy\right],$}
$$ 
we derive
$$
\adjustbox{width=4.513in}{$\displaystyle \bigg|\left(-\frac{1}{2}P\mathcal{D}^{(2)}_{\varphi}P\right)_{i,j}
-\kappa^m_{c}(x_i,x_j)\bigg| = \frac{1}{2}\bigg|-R^{(1)}
-R^{(2)}
+R^{(3)}
+R^{(2)}\bigg|\sim O\left(\frac{1}{m}\right)
\quad \forall x_i,x_j \in [0,1]. $}
$$
This completes the proof.
\end{proof}

\subsection{Proof of Theorem \ref{thm:changepoint eigenfunction}}
\begin{proof}
Consider a function $f(y)= g(y)$ for $y\leq t^*$, $f(y)=h(y)$ for $y>t^*$, where $g(t^*)=h(t^*)$. If $f$ is an eigenvector of the integral operator induced by $\kappa^m_{c}$, then
\begin{align*}
&-\frac{1}{2}\int_0^{1} \mathcal{DC}[\kappa^m(x,y)] f(y)\,\mathrm{d}y= \lambda f(x),\text{ so}\\
(x<t^*):&\quad \int_0^{t^*} \mathcal{DC}[\kappa^m(x,y)] g(y)\,\mathrm{d}y+\int_{t^*}^1\mathcal{DC}[\kappa^m(x,y)] h(y)\,\mathrm{d}y= -2\lambda g(x)\\
(x>t^*):&\quad \int_0^{t^*} \mathcal{DC}[\kappa^m(x,y)]g(y)\,\mathrm{d}y+\int_{t^*}^1 \mathcal{DC}[\kappa^m(x,y)] h(y)\,\mathrm{d}y= -2\lambda h(x).
\end{align*}
We compute the second derivative with respect to $x$ for each of the two cases; specifically $x<t^*$ and $x>t^*$:
\begin{align*}
(x < t^*):\quad &\frac{\partial^2}{\partial x^2} \int_0^{t^*} \mathcal{DC}[\kappa^m(x,y)] g(y) \, \mathrm{d}y = \int_0^{t^*} \frac{p\!-\!p^2}{m} (-2) g(y) \, \mathrm{d}y\\
&  + \frac{\partial^2}{\partial x^2} \int_0^x \frac{p\!-\!p^2}{m} (x\!-\!y) g(y) \, \mathrm{d}y + \int_x^{t^*} \frac{p\!-\!p^2}{m} (y\!-\!x) g(y) \, \mathrm{d}y \\
&\adjustbox{width=4in}{$\displaystyle = \frac{p\!-\!p^2}{m} (-2)[G(t^*) \!-\! G(0)] + \frac{\partial}{\partial x} \int_0^x \frac{p\!-\!p^2}{m} g(y) \, \mathrm{d}y + \int_x^{t^*} \frac{p\!-\!p^2}{m} (-1) g(y) \, \mathrm{d}y$} \\
&= \frac{p\!-\!p^2}{m} \left[(-2)(G(t^*) \!-\! G(0)) \!+\! 2g(x)\right].\\
(x < t^*):\quad &\frac{\partial^2}{\partial x^2} \int_{t^*}^1 \mathcal{DC}[\kappa^m(x,y)] h(y) \, \mathrm{d}y \\
&= \int_{t^*}^1 \frac{p\!-\!p^2}{m} (-2) h(y) \, \mathrm{d}y \\
&= \frac{p\!-\!p^2}{m} (-2) [H(1) \!-\! H(t^*)].
\end{align*}

We observe that $G(t^*)-G(0)+H(1)-H(t^*)=\int_0^{t^*} g(y)\,\mathrm{d}y+\int_{t^*}^1 h(y)\,\mathrm{d}y= \int_0^1 f(y)\,\mathrm{d}y=0,$ so the first equation (for $x<t^*$) becomes
$$\frac{2p(1-p)}{m} g(x) = -2\lambda g''(x).$$

For the second equation,
\begin{align*}
(x>t^*):\quad \frac{\partial^2}{\partial x^2} \int_0^{t^*} \mathcal{DC}[\kappa^m(x,y)] g(y) \, \mathrm{d}y&= \int_0^{t^*} \frac{q-q^2}{m} (-2) g(y)\,\mathrm{d}y,\\
&= \frac{q-q^2}{m}(-2)[G(t^*)-G(0)].
\end{align*}
\begin{align*}
(x>t^*):\quad &\frac{\partial^2}{\partial x^2} \int_{t^*}^{1} \mathcal{DC}[\kappa^m(x,y)] h(y)\,\mathrm{d}y\\
&= \int_{t^*}^1 \frac{q-q^2}{m} (-2) h(y)\,\mathrm{d}y\\
&\quad+\frac{\partial^2}{\partial x^2} \int_{t^*}^x \frac{q-q^2}{m} (x-y) h(y)\,\mathrm{d}y+ \int_{x}^1 \frac{q-q^2}{m} (y-x) h(y)\,\mathrm{d}y\\
&= \frac{q-q^2}{m} (-2) [H(1)-H(t^*)]\\
&\quad +\frac{\partial}{\partial x} \int_{t^*}^x \frac{q-q^2}{m} h(y)\,\mathrm{d}y+\int_x^1 \frac{q-q^2}{m} (-1) h(y)\,\mathrm{d}y\\
&= \frac{q-q^2}{m} \left[(-2)(H(1)-H(t^*))+2h(x)\right].
\end{align*}

Again applying the condition $\int_0^1 f(y)\,\mathrm{d}y=0$, the equation for $x>t^*$ reads $$ \frac{2q(1-q)}{m} h(x)=-2\lambda h''(x). $$

Assuming $\lambda>0$, the solutions to the differential equation for $g$ must have the form $e^{rx}$ where $r$ is a root of the equation $r^2+\frac{p(1-p)}{m\lambda}=0$. In other words, $r= \pm i\sqrt{\frac{p(1-p)}{m\lambda}}$ for $\lambda>0$. Arguing similarly for $h$, we see that the solutions must have the form
\begin{align*}
g(x)&= A_g\cos\left(\sqrt{\frac{p(1-p)}{m\lambda}}x\right)+B_g\sin\left(\sqrt{\frac{p(1-p)}{m\lambda}}x\right)\\
h(x)&= A_h\cos\left(\sqrt{\frac{q(1-q)}{m\lambda}}x\right)+B_h\sin\left(\sqrt{\frac{q(1-q)}{m\lambda}}x\right).
\end{align*}
\end{proof}

\section{Summary statistics over time for organoid networks}
\label{sec:organoiddetails}

We analyze the time series of 30 approximate functional connectivity graphs derived from brain organoid recordings, where all graphs in the series are directed and unweighted, then characterize the network structure using three fundamental summary statistics: edge density, average path length, and reciprocity.

When $\mathbf{A}$ is the adjacency matrix for an unweighted and directed graph with $n$ nodes, edge density is defined as the ratio of the number of observed edges to the number of total possible edges:
$$\text{Edge Density} (\rho) := \frac{ \mathbf{1}^\top \mathbf{A} \mathbf{1} }{ n(n-1)},$$where $\mathbf{1}$ is an all-ones vector of length $n$.

Average path length (APL) is defined as the mean of the shortest directed path distances between all pairs of nodes where a connection exists.
In the directed network, it accounts for both directions ($A \to B$ and $B \to A$). 
When the graph is strongly connected, meaning one can get from every node to every other node by strictly following the direction of the arrows, APL can be calculated by:
$$\text{Average Path Length} (L) := \frac{ \mathbf{1}^\top \mathbf{D} \mathbf{1} }{ n(n-1) },
$$
where $\mathbf{D}$ is the matrix of shortest directed path lengths.

Reciprocity measures the tendency of node pairs to form mutual connections. 
It is calculated as:
$$\text{Reciprocity} := \frac{ \mathbf{1}^\top \left(\mathbf{A}^{\top} \circ \mathbf{A} \right) \mathbf{1} }{ \mathbf{1}^\top \mathbf{A} \mathbf{1} },$$
where $\circ$ is the entrywise matrix product.

\begin{figure}
    \centering
    \includegraphics[width=1\linewidth]{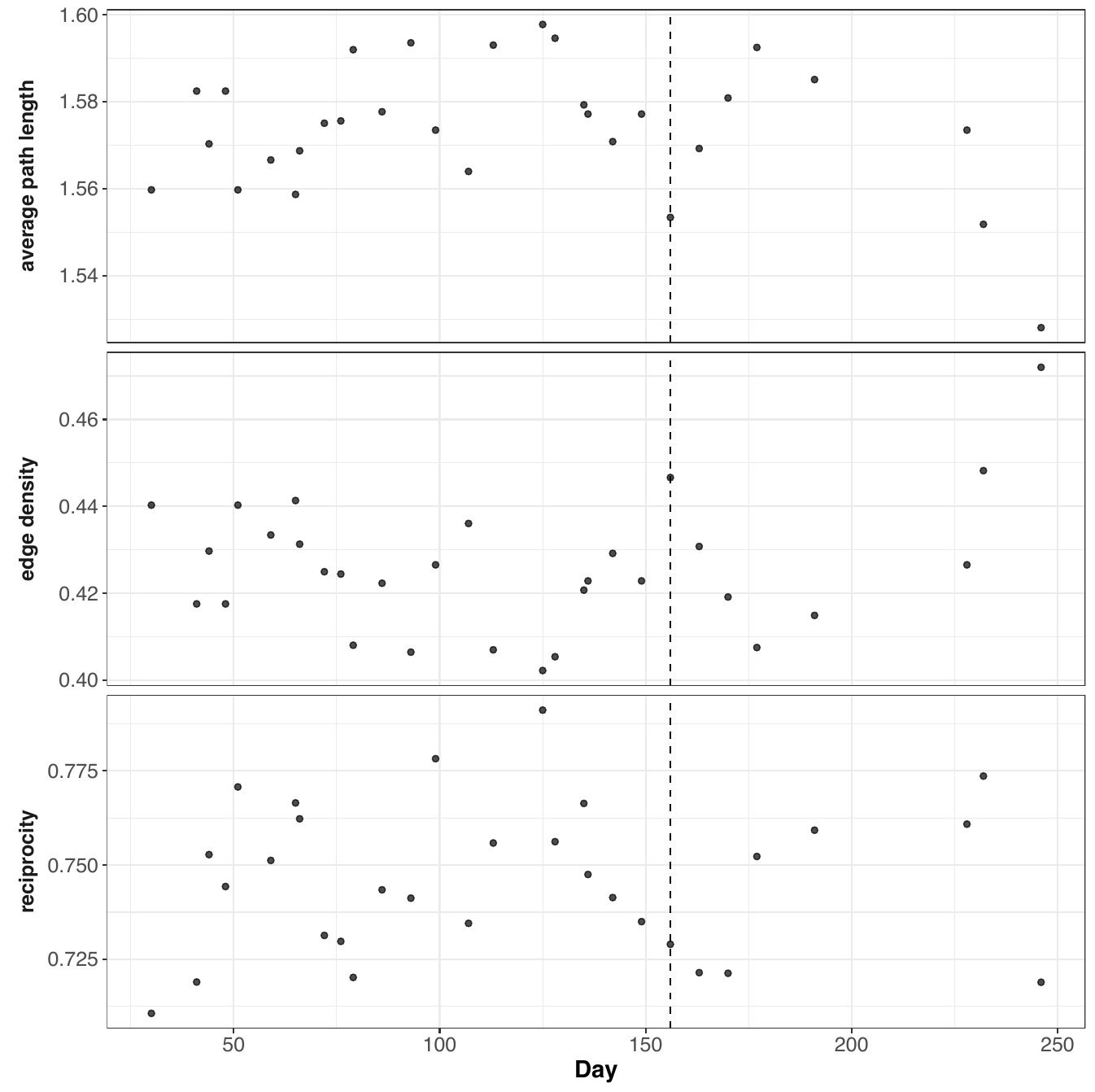}
    \caption{Summary statistics varying over time for brain organoid data.}
    \label{fig:Well34_summary}
\end{figure}

The results reveal a structural property of these functional networks:
as shown in Figure \ref{fig:Well34_summary}, there is a strict linear relationship between average path length and edge density.
Specifically, we can confirm that for every graph in the series:
$$\text{Edge Density} + \text{Average Path Length} = 2.$$
We can also confirm that every graph is strongly connected because we choose the largest common connected component. For these graphs, the maximum path length is 2.
In other words, if a direct edge does not exist between any two nodes $i$ and $j$, there exists a distinct intermediate node $k$ such that $i \to k \to j$. This implies that the distance matrix $\mathbf{D}$ has all entries equal to 1 or 2.
Let $M = \mathbf{1}^\top \mathbf{A} \mathbf{1}$ be the total number of edges. 
The total number of node pairs is $T = n(n-1)$.
The sum of all path lengths is:
$$ \mathbf{1}^\top  \bD \mathbf{1} = 1 \times M + 2 \times (T - M) = 2T - M.$$
Dividing by $T$ to find the average ($L$):$$L = \frac{2T - M}{T} = 2 - \frac{M}{T} = 2 - \rho$$
Thus $\rho + L = 2$.

This structural constraint indicates that the functional networks are highly integrated, with information flowing rapidly across the entire graph. 
The high reciprocity ($\approx 0.72$) further supports this, suggesting a network dominated by mutual feedback loops.

However, despite this constant structural diameter, the specific connectivity fluctuates significantly over time. 
The constant flux observed in the summary statistics is similar to the behavior seen in the control charts, indicating that the underlying network dynamics are continuously evolving and cannot be described as a simple zeroth-order changepoint.
We also note that these networks are dissimilar to the Model \ref{changepointmodel}, since edge densities are significantly fluctuating over time, while in the Model \ref{changepointmodel} we would expect the edge density to monotonically increase. In particular, none of these marginal statistics demonstrate a clear changepoint at the appropriate time.

\end{appendix}

\section*{Funding}
This work was supported by National Science Foundation (SES-1951005), Office of Naval Research (ONR) Award Number N00024-22-D-6404 (through JHU Applied Physics Laboratory), Office of Naval Research SoA: N00014-24-1-2278, the Amazon-JHU AI2AI Initiative, Microsoft Research, the Acheson J. Duncan Fund for the Advancement of Research in Statistics.

\bibliographystyle{unsrtnat} %
\bibliography{references2}       %

\end{document}